\documentstyle[aps,multicol,prb,epsfig]{revtex}

\begin{document}

\draft

\title{Conductance fluctuations as a tool for investigating the quantum modes
\newline in atomic size metallic contacts}
\author{B. Ludoph and J.M van Ruitenbeek}
\address{Kamerlingh Onnes Laboratorium, Universiteit Leiden, Postbus 9504,
2300 RA Leiden, The Netherlands}

\maketitle

\begin{abstract}
Recently it has been observed that the conductance fluctuations of atomic size
gold contacts are suppressed when the conductance is equal to an integer multiple of the conductance quantum. The fact that these contacts tend to consist exclusively of fully open or closed modes has been argued to be the origin for this suppression. Here, the experiments have been extended to a wide range of
metallic elements with different chemical valence and they provide new
information about the relation between the mode composition
and statistically preferred conductance values observed in conductance
histograms.
\end{abstract}


\begin{multicols}{2}
\settowidth{\columnwidth}{aaaaaaaaaaaaaaaaaaaaaaaaaaaaaaaaaaaaaaaaaaaaaaaaa}

\section*{Introduction}
Manipulation and characterization of atoms and atomic size metallic
constrictions has recently become available through the development of the STM
\cite{stm1atom}. An alternative tool, for creating stable and clean atomic
size metallic contacts, is the mechanically controllable break junction (MCB)
\cite{chris,krans1}. For the characterization of these systems, measurements of
the electrical conductance are widely employed. This, as a result of the ease with which they
usually can be obtained. The framework within which one should describe the
conductance of such small contacts, which in the case of metals have dimensions
of order of the Fermi wavelength, is the Landauer-B\"{u}ttiker formalism
\cite{landauer}. In this formalism the conductance in the contact is described
by $N$ channels, determined by the narrowest cross-section of the constriction
and the Fermi wavelength. Each channel has a transmission probability $T_{n}$
with a value between 0 and 1. The total conductance is given by $G =
(2e^{2}/h)\sum_{n=1}^{N} T_{n}$. For an adiabatic constriction in a free
electron gas, the conductance increases stepwise with quantum units of the
conductance ($G_{0} = 2e^{2}/h$) as the channels open one by one while
increasing the constriction diameter \cite{2deg}. However, when one pulls apart
a metallic atomic size contact, neither the diameter nor the conductance of the
constriction decreases smoothly. Instead, a series of steps (of order $G_{0}$)
and plateaus are observed in the conductance on elongation of the contact. The
sequence of steps and plateaus is different each time the contact is pulled
apart. The steps correspond with atomic reconfigurations and the plateaus with
elastic deformation of the contact \cite{agrait}. It is tempting, but in
principle incorrect, to assume offhand that these conductance measurements on
atomic necks simply probe a series of discrete diameters of a free electron gas.
It may work for some metals (we will show that for sodium this is nearly the
case), but a correct general description of the conductance of metallic point
contacts consisting of even in the simplest case, a single atom, has to consider
the chemical valence of this atom \cite{cuevas,scheer2}. Compare for instance a
single atom contact of the monovalent $s$-metal gold with the trivalent $sp$-
metal aluminum. In the former case the conductance is carried by a single
channel with conductance close to $G_{0}$. A single atom of aluminum on the
other hand, with three conduction electrons in the $3s$ and $3p$ shells, also
has a total conductance close to $G_{0}$ but allows three partially transmitting
channels.

Recently \cite{me} a new technique, making use of conductance fluctuations, has
been presented which does not require superconductivity \cite{scheer2,scheer} to
obtain information about the conductance modes contributing to the conductance.
First results on gold contacts with conductance up to 5\,$G_{0}$ have shown that
for this $s$-metal, once a channel (with number $n$) is partially open it tends
to
fully open before a next $(n+1)^{\text{th}}$ channel starts to contribute
significantly. This
interpretation was confirmed by an independent technique which consists of
measuring the shot noise in the point contact current \cite{vdBrom}. In this
paper we present a more complete argumentation of the theory together with new
measurements of the conductance fluctuations on copper, silver, sodium, aluminum,
niobium and iron and discuss what information on the channel transmissions can
be extracted from our results. We will also compare our conductance fluctuation
results to recently published measurements on the thermopower of atomic size
contacts and show that both measurements can be related without any free
parameters.

\begin{figure}[htb]
\begin{center}
\leavevmode
\epsfig{file=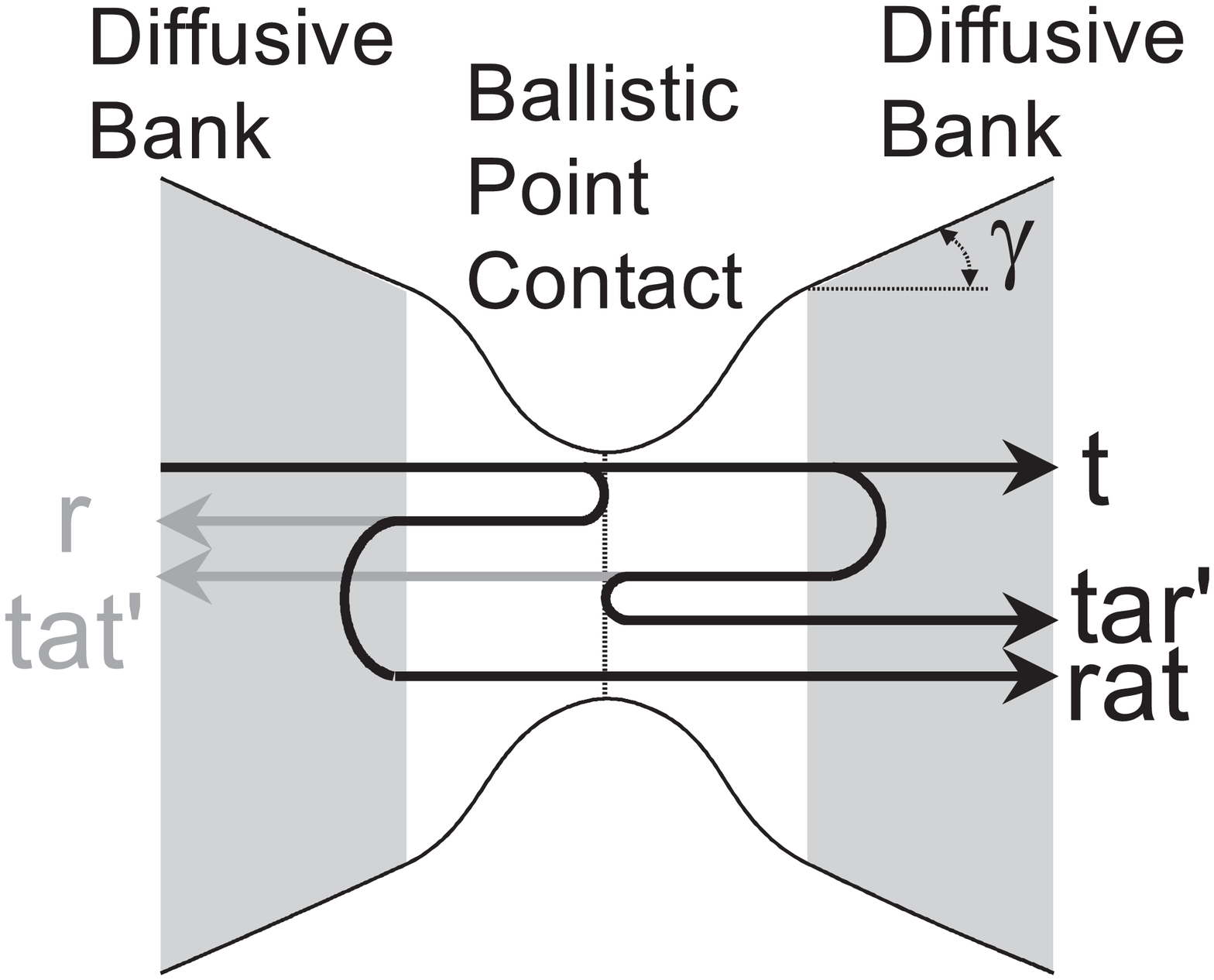,width=6cm} \caption{Schematic
diagram of the configuration used in the theoretical analysis. The
dark lines with arrows show the paths, which interfere and
contribute to the conductance fluctuations in lowest order.}
\end{center}
\label{figschemat}
\end{figure}

Fluctuations in the conductance with bias voltage have previously been observed
in larger ballistic contacts \cite{holweg} and have an origin analogous to the
universal conductance fluctuations (UCF) measured in diffusive wires \cite{ucf}. The
interesting new aspect of such fluctuations in quantum point contacts is that
their r.m.s. amplitude depends on the transmission probability of the channels
contributing to the conductance. The underlying principle of this effect can be
understood by considering a contact with a single conducting mode having a
finite transmission probability $T$, described by transmission and reflection
coefficients $t$, $t^{\prime}$, $r$ and $r^{\prime}$ (coming in from left and
right, respectively), with $|t^{\prime}|^{2}=|t|^2=T$ and
$|r^{\prime}|^{2}=|r|^2=1-T$. As illustrated in Fig.\,\ref{figschemat}, electron
waves transmitted by the contact with amplitude $t$, and scattered back towards
the contact through diffusive paths in the bank with probability amplitude $a$,
have a probability amplitude $r$ to be {\it reflected} at the contact. This wave
interferes with the directly transmitted partial wave and modifies the total
conductance, depending on whether the resulting interference is constructive or
destructive. A similar contribution comes from the trajectories on the other
side of the contact. The interference terms will be sensitive to changes in
the phase accumulated along the trajectories, which is determined by the
electron energy and the path length. The fluctuations in the conductance as a
function of bias voltage are thus the result of the change of these phase
factors by the increase in  the kinetic energy of the electrons by an amount
$eV$. What is immediately apparent from the principle illustrated in
Fig.\,\ref{figschemat} is that when the coefficient $T$ is either 0, or 1, the
interference and thus the amplitude of the fluctuations vanishes. This
suppression of conductance fluctuations at quantized values has been noted in
numerical simulations of quantum point contacts containing disorder by Maslov
{\it et al.} \cite{maslov}

Each time when the contact is opened, and closed again to sufficiently large
conductance values, random atomic reconfigurations take place, leading to a
completely new set of scattering centers. The statistical results of many
different contacts can hence be interpreted as the ensemble average over defect
configurations. With this technique we have studied the average properties of
the conductance modes for different materials and their relation with the
statistically preferred conductance values observed previously by various
authors \cite{krans,brand,gai,costa,sirvent,yanson} through the measurement of
so called conductance histograms.

Typically when studying conductance fluctuations one measures the differential
conductance over a wide range of bias voltage or magnetic field. Here, on the
other hand, we measure the first and second derivative of the current with
respect to voltage of atomic size contacts. The first derivative gives us the
conductance, and the second derivative is a parametric derivative of the
conductance. The latter can roughly be seen as a measure for the amplitude of the
fluctuations with voltage. We measure these quantities for a large number of different
contacts, thus determining effectively an average over an ensemble of scattering
configurations. This measurement method is preferable as it is much faster than
measuring the conductance as a function of bias voltage directly, and hence
allows the experimental determination of the average properties of many contacts
within a reasonable time scale.

\section*{Theory}

In this section we will give a more detailed description of the phenomenological
theory presented in Ref.\,\onlinecite{me}. In our model for a metallic
constriction we
divide the conductor into three separate regions: A region, small on the scale
of the scattering lengths involved, centered around the narrowest part of the
conductor, which we describe as ballistic. On either side of this ballistic
region we consider a diffusive region, characterized by a mean free path,
$l_{e}$. In order to make the geometry of the contact more realistic, we assume
a conical shape for these diffusive regions, with opening angle $\gamma$ (see
Fig.\,\ref{figschemat}). The probability amplitudes for scattering from any
incoming mode on the left to any outgoing mode on the right side of this
ballistic section (henceforth referred to as the 'bare contact') is described in
terms of the matrices of transmission $\bf{t}$, $\bf{t}^{\prime}$ and reflection
$\bf{r}$, $\bf{r}^{\prime}$ when coming from the left and right respectively.

\begin{eqnarray}
\left(
\begin{array}{l}
o_{r} \\ i_{r}
\end{array}
\right) &=& {\bf M}\times \left(
\begin{array}{l}
i_{l} \\ o_{l}
\end{array}
\right) \nonumber \\ &=& \left(
\matrix{{\bf t}^{\dagger -1} & {\bf r}^{\prime }{\bf t}^{\prime -1}\cr
-{\bf t}^{\prime -1}{\bf r} & {\bf t}^{\prime -1}\cr}
\right) \left(
\begin{array}{l}
i_{l} \\ o_{l}
\end{array}
\right), \label{mscat}
\end{eqnarray}

\noindent where ${\bf M}$ is the transfer matrix and $i_{r}$, $o_{r}$,
$i_{l}$, $o_{l}$ are
the vectors of the incoming and outgoing waves for the right and left hand side
respectively.
The matrix of transmission probabilities is given by ${\bf
T}=\bf{tt}^{\dagger}$, which can be diagonalized \cite{butt,mads}. For a narrow
constriction, most of the diagonal elements will be zero. The number of
conducting modes
$N$ and their transmission probability is given, by the
non-zero diagonal elements, $ T_{n}=|t_{n}^{2}|=|t_{n}^{\prime 2}|$ $n
=1,2,...,N$. The reflection probability of mode $n$ is given by,
$R_{n}=|r^{2}_{n}|=|r^{\prime 2}_{n}|=1-T_{n}$.
In the simple free electron gas model, the number $N$ is determined by the width of the
narrowest part of the contact \cite{landauer} and by the Fermi wavelength. Note, however, that in principle it is not restricted to any particular model.
The
values of the $T_{n}$'s are somewhat influenced by our arbitrary choice of the
boundaries between the ballistic and diffusive regions. For this influence to be
small, the distance $L$ between the center of the contact and the boundaries
should be large on the scale of the contact diameter. On the other hand, in
order to be able to neglect fluctuations of $T_{n}$ on the scale of the applied
voltage, $V$, we require  $L \ll \hbar v_{F}/eV$. For metallic contacts we can
typically take $L \simeq 1$\,nm.

The left and right banks are also described in terms of transfer matrices
${\bf M}_{l,r}$, similar in form to the one used for the bare
contact in Eq.\,\ref{mscat}. We define in this case ${\bf t}_{r}$ and ${\bf
a}_{r}$ as the transmission matrix and return amplitude matrix for the right
bank and ${\bf t}_{l}$ and ${\bf a}_{l}$ the corresponding ones for the left
bank. The elements of the return amplitude matrices $a_{l_{mn}}$ and
$a_{r_{mn}}$, which scatter a wave from mode $m$ to mode $n$ in the left or
right hand side diffusive section, are expected to be small compared to 1. It is
hence, a reasonable approximation to calculate the total transmission
probability to first order in the return amplitude matrix elements only. The
return amplitudes are energy dependent but this will not be made explicit until
this dependence becomes relevant in Eq.\,\ref{deltag}. The total transmission
matrix for the two banks and ballistic constriction combined can be
written as,

\begin{eqnarray}
{\bf t}_{t}&=&(({\bf M}_{r}{\bf MM}_{l})_{22})^{-1}  \label{tt} \\
&=& {\bf t}_{l}^{\prime}\left( {\bf t}^{\prime -1} - {\bf a}_{r}{\bf t}^{\dagger
-1}{\bf a}_{l} - {\bf t}^{\prime -1}{\bf ra}_{l}-{\bf a}_{r}{\bf r}^{\prime}{\bf
t}^{\prime -1} \right)^{-1}{\bf t}^{\prime}_{r}. \nonumber
\end{eqnarray}

\noindent Since the return amplitudes will usually be small we can set
${\bf t}_{l} = {\bf t}_{r} \simeq {\bf I}$, the identity matrix. With this
assumption and the fact that we are only calculating to lowest order in
$a_{l,r_{mn}}$ we are neglecting corrections to the total conductance and higher
order contributions to the conductance fluctuations. These corrections will be
discussed at the end of this section.

In order to calculate the conductance fluctuations, we will make an expansion
of the total transmission probability to lowest order in $a_{l,r_{mn}}$. Using $
{\bf t}_{t}{\bf t}_{t}^{\dagger }  = ((
{\bf t}_{t}^{-1}) ^{\dagger }({\bf  t}_{t}^{-1})) ^{-1}$
and assuming the matrices ${\bf t}$, ${\bf t}^{\prime}$, ${\bf r}$ and ${\bf
r}^{\prime}$ are already in diagonal form, the trace of ${\bf t}_{t}{\bf
t}_{t}^{\dagger }$ to lowest order in $a_{l,r_{mn}}$ can be approximated by the
sum of the inverse of the diagonal components of $(
{\bf t}_{t}^{-1}) ^{\dagger }({\bf  t}_{t}^{-1})$,

\begin{eqnarray}
\text{Tr}[{\bf t}_{t}{\bf t}_{t}^{\dagger }]&\approx& \sum_{n=1}^{N}\left(
\frac{1}{
T_{n}^{-1}-T_{n}^{-1}2\text{Re}(a_{r_{nn}}r_{n}^{
\prime }+r_{n}a_{l_{nn}})}\right) \nonumber \\
&\approx& \sum_{n=1}^{N}T_{n}\left( 1+2
\text{Re}(a_{r_{nn}}r_{n}^{\prime }+r_{n}a_{l_{nn}})\right). \label{trttd}
\end{eqnarray}

\noindent We assume that the Boltzmann constant times the temperature,
$k_{B}\theta$, is much smaller than
the energy scale of the applied voltage, $eV$, (in accordance with the situation
in our experiment) so that we can take the zero temperature approximation. The
current is then determined by Eq.\,\ref{trttd} through \cite{landauer},

\begin{equation}
I= \frac{2e^{2}}{h}\int_{0}^{eV} \text{Tr}[{\bf t}_{t}{\bf t}_{t}^{\dagger}]dE.
\label{eqI}
\end{equation}

\noindent The fluctuations in the conductance are described by $\delta G = G -
\langle G \rangle$, with $G=\partial I/\partial V$. Combining this with
Eq.\,\ref{trttd} and Eq.\,\ref{eqI}, and using $\langle G \rangle =
(2e^{2}/h)\sum_{n=1}^{N}T_{n}$ to first order in $a_{l,r_{mn}}$, leads to,

\begin{eqnarray}
\delta G(V) &=& \sum_{n=1}^{N}
\frac{\partial}{\partial eV}\int_{0}^{eV}\frac{2e^{2}}{h}T_{n} \label{deltag} \\
& &\times 2 \text{Re}[r_{n}a_{l_{nn}}(eV-E) + a_{r_{nn}}(-E)r_{n}^{\prime }]dE
\nonumber \\
&=& -\sum_{n=1}^{N} \frac{2e^{2}}{h} T_{n} 2 \text{Re}[r_{n}^{\prime
}a_{r_{nn}}(eV)-r_{n}a_{l_{nn}}(-eV)]. \nonumber
\end{eqnarray}

\noindent The return amplitudes contain random phase factors of the form exp$[-i(E_{F} \pm
eV)\tau/\hbar]$, where $\tau$ is the traversal time for a particular trajectory.
Averaging Eq.\,\ref{deltag} over the ensemble of defect configurations will give
a zero result, $\langle \delta G \rangle = 0$. For the correlation function of
the conductance as a function of voltage, however, we obtain a finite
contribution. In the product, only terms of the form
$a_{l,r_{nn}}(E_{1})a^{*}_{l,r_{nn}}(E_{2})$ have a chance to survive the
averaging. In addition, diffusion in the left and right banks is uncorrelated,
so that products of $a_{l_{nn}}(E_{1})$ and $a_{r_{nn}}(E_{2})$ average to zero,
\begin{eqnarray}
& &\left\langle \delta G(eV_{1})\delta G(eV_{2})\right\rangle
= \sum_{n=1}^{N}\left( \frac{2e^{2}}{h}T_{n}\right)^{2}R_{n} \label {eqv1v2}\\
&\times& \left\langle
2\text{Re}[a_{r_{nn}}(eV_{1})a_{r_{nn}}^{*}(eV_{2})
+ a_{l_{nn}}(-eV_{1})a_{l_{nn}}^{*}(-eV_{2})]\right\rangle.
\nonumber
\end{eqnarray}

\noindent At this point we assume that the average properties of the scattering
on both sides of the contact, for all the mode indexes, are the same. Further,
we propose that $\langle a_{l,r_{nn}}(E_{1})a_{l,r_{nn}}^{*}(E_{2})\rangle$ can
be expressed as $\int_{0}^{\infty} P_{cl}(\tau) e^{-i(E_{1}-E_{2})\tau/\hbar}
\text{d}\tau $, where $\tau$ is the time required for the completion of a
classical diffusive trajectory, and $P_{cl}(\tau)$ is the classical probability
distribution to return to the contact at this time. The classical return
probability, $P_{cl}(\tau)$, can be obtained by considering an electron being
injected from the ballistic central section of the contact into the diffusive
region at the left or right. When we take the interface between the ballistic
and diffusive regions to be at a small distance $L$ from the contact center,
then after a given time $\tau$, with $D\tau \gg L^{2}$, the probability
distribution to find the electron at a distance $r > L$ is given by the
classical result,

\[ \rho(\tau,r) d\tau = \frac{2}{(1-\text{cos}\gamma)(4 \pi D
\tau)^{3/2}}\text{e}^{(r-L)^{2}/4D\tau}d\tau,
\]

\noindent where $D=v_{F}l_{e}/3$ is the diffusion constant with $l_{e}$ the mean
free path for elastic scattering. Here we have assumed that the diffusive region
has the shape of a cone with opening angle $\gamma$ (Fig.\,\ref{figschemat}) and
that only a small number of channels are transmitting, so that most electrons
entering the ballistic region are reflected. The probability per unit time to
find the particle back in a disk of radius $\sigma$ and thickness $dx$ at the
entrance of the ballistic region is $\rho(\tau,L) \pi \sigma^{2} dx$. The
average time the particle spends in this volume is $dx/\langle v_{x} \rangle =
\sqrt{3} dx/v_{F}$. The probability that the particle moves towards the contact
instead of away from it is 1/2, and we assume that it has equal probability to
enter into any of the $N_{b}$ modes available at the entrance of the ballistic
section, where $N_{b} = (k_{F} \sigma/2)^{2}$. Thus, the probability per unit
time to return to the contact after a time $\tau$, into a given mode, is

\begin{eqnarray} P_{cl}(\tau) = \frac{v_{F}}{2 \sqrt{3 \pi} k_{F}^{2} (D
\tau)^{3/2} (1-\text{cos}\gamma)}. \label{pcl}
\end{eqnarray}

\noindent This distribution should further be multiplied with a factor
$\text{e}^{\tau/\tau_{\phi}}$ in order to take into account the probability that
within a typical time $\tau_{\phi}$ the electron undergoes scattering which
destroys phase memory. Combining these expressions we obtain,
\begin{eqnarray}
\langle \delta G(V_{1}) \delta G(V_{2}) \rangle = \sum_{n=0}^{N} 4\left(
\frac{2e^{2}}{h} T_{n} \right)^{2} (1-T_{n}) \nonumber \\
\times \int_{0}^{\infty} P_{cl}(\tau) \text{cos} \frac{e(V_{1} - V_{2})
\tau}{\hbar} e^{-\tau/\tau_{\phi}} d \tau.
\label{deltagdeltag}
\end{eqnarray}
In our measurements we really measure $\langle (\partial G /\partial
V)^{2}\rangle$ with a fixed modulation voltage rather than $\langle \delta
G(eV_{1})\delta G(eV_{2})\rangle$. This can easily be corrected by
differentiating Eq.\,\ref{deltagdeltag} with respect to $V_{1}$ and $V_{2}$, and
then setting $V=V_{1} = V_{2}$. In the limit of $V \rightarrow 0$ and using the
above approximation for the average return probability we obtain,

\begin{eqnarray*}
\sigma _{GV}^{2} &\equiv&\left\langle \left( \frac{\partial \delta G}{\partial
V}\right)
^{2}\right\rangle =
G_{0}^{2}\sum_{n=0}^{N}T_{n}^{2}(1-T_{n})\nonumber \\
& & \times \frac{4e^{2}v_{F}}{\sqrt{
3\pi }\hbar ^{2}k_{F}^{2}}\frac{1}{1-\text{cos}\gamma}
\frac{1}{D^{3/2}}\int_{0}^{\infty }\sqrt{\tau }\text{e}^{-\tau/\tau_{\phi}}
d\tau .
\end{eqnarray*}

\noindent Evaluating the integral and taking the square root, results in,
\[
\sigma _{GV}= \frac{\sqrt{6}e G_{0}}{\hbar k_{F}v_{F} \sqrt{1-
\text{cos}\gamma}}\left(
\frac{\tau_{\phi }}{\tau_{e}}\right) ^{3/4}\sqrt{\sum_{n=0}^{N}T_{n}^{2}(1-
T_{n})}.
\]

\noindent The typical time scale on which a first collision takes place is the
elastic scattering time $\tau_{e} = l_{e}/v_{F}$.

We have conducted our experiment by measuring the first and second derivative of
the current with respect to voltage. The amplitude of the applied modulation
voltage was responsible for the energy cut-off rather than the dephasing time as
we have assumed above ($eV_{mod} > \hbar/\tau_{\phi}$). Using the derivation in
the Appendix, which incorporates this finite modulation voltage into the theory
gives us the final result \cite{correction},

\begin{equation}
\sigma_{GV}=\frac{2.71 e G_{0}}{\hbar k_{F}v_{F} \sqrt{1-
\text{cos}}\gamma}\left(
\frac{
\hbar /\tau_{e}}{eV_{mod}}\right)^{3/4}\sqrt{\sum_{n=1}^{N}T_{n}^2(1-T_{n})}.
\label{sigma}
\end{equation}

For the fluctuations in the conductance we have described above, terms higher in
order than $a_{l,r_{mn}}$, were not very important. However, when all the $N$
channels contributing to the conductance are fully open, the first
order contribution we have calculated above will be zero. Under this
condition the second order terms may have a noticeable contribution. In
this case we can take ${\bf t} = {\bf I}$ and ${\bf r} = {\bf 0}$ in Eq.\,\ref{tt} which
greatly
simplifies the derivation. It is then quite easy to show that at quantized
values $G=NG_{0}$, the contribution of the second order term in $a_{l,r_{mn}}$ is
$\sigma_{GV} \propto N\sqrt{ \langle |a_{l_{mn}}|^{2}|a_{r_{mn}}|^{2}\rangle}$.
These terms are too small to explain the reduction of the depths of the minima
in the experiment discussed below and will be further ignored.

These higher order terms, however, are not negligible when we consider the total
conductance of the contact. The importance of these higher order corrections
becomes apparent when we compare our theory to the experiments, and notice that
these effects result in a significant correction to the total conductance $G$.
The necessity to include these conductance corrections in our
analysis is a direct consequence of the fact that in the experiment we can not
measure the bare contact conductance alone, as we always measure it in series
with the banks. This feature in the conductance of quantum point contacts is
usually referred to as the series resistance \cite{krans,brand,pgarcia}.

The lowest order correction to the average total transmission probability is
given by
$\langle \text{Tr}[{\bf t}_{t}{\bf t}_{t}^{\dagger}]\rangle = \sum_{n=1}^{N}
T_{n} (1-
\sum_{m=1}^{N}T_{m}(\langle|a_{l_{mn}}|^{2}\rangle +
\langle|a_{r_{mn}}|^{2}\rangle))$. The last
term describes the path of an electron that is transmitted through the contact,
scattered back towards the contact in the diffusive bank, and then transmitted
through the contact a second time in opposite direction. These processes will
lead to a smaller conductance than expected for the bare contact conductance
alone since part of the transmitted electrons are scattered back, reducing the
net forward current flow.

At higher conductance values, we expect a significant
contribution of even higher order terms in the return probability
$a_{l,r_{mn}}$, to the
conductance correction, and that hence the lowest order correction used above
will not suffice.
Keeping track of higher order terms, becomes very
complicated for many channels. However, using random matrix theory an expression
for the correction to the conductance of a quantum point contact connected to
diffusive leads has already been derived by Beenakker and Melsen \cite{melsen},

\begin{eqnarray}
\frac{\langle G \rangle}{G_{0}} =
\frac{g}{1+(g+1)r}+\frac{1}{3}\left( \frac{(g+1)r}{1+(g+1)r}\right)^{3}.
\label{rcorr}
\end{eqnarray}

\noindent Here $g=\sum_{n=1}^{N} T_{n}$ is the reduced conductance of the
bare contact, where in the theory all channels were assumed to be perfectly
transmitting. The diffusive scattering in the banks is represented by
$r=G_{0}/G_{d}$, with $G_{d}$ the diffusive conductance of the banks.
When one makes the assumption that the conductance of the banks is large
compared to
the conductance of the contact, Eq.\,\ref{rcorr} can be simplified and rewritten
to a form where the correction to the conductance effectively becomes a somewhat
contact  dependent series resistance,
\begin{eqnarray}
\langle \delta R(g) \rangle \approx
\left(1+\frac{1}{g}\right)\left(\frac{1}{G_{d}}\right).
\label{rcorrsimp}
\end{eqnarray}

\noindent To lowest order Eqs.\,\ref{rcorr} and \ref{rcorrsimp} are consistent
with the correction to the average total transmission probability derived from
the backscattering above.

\section*{Experimental Method}

We have used an MCB to make stable atomic size metallic contacts. The technique
\cite{chris,krans1}, uses a notched metal wire glued onto an electrically
insulated phosphor
bronze bending beam. In the case of sodium a slightly different sample fixation
method was required due to its high oxidation rate (see Krans {\it et al.}
\cite{krans}). The sample is placed in a three point bending configuration,
which consists of two counter supports and a drive at the center of the bending
beam powered by a piezoelectric element in combination with a mechanical screw.
By first turning the screw and later expanding the piezo by applying a voltage
over it, we can bend the substrate in a controlled way elongating the wire until
it finally breaks. The wire is broken at low temperature (4.2\,K) in an
evacuated can to ensure that two clean freshly broken surfaces are measured. The
voltage applied over the piezoelectric element is linear with the elongation of
the contact (for further details see Ref\,\onlinecite{curacau}).

The conductance measurements are performed by applying a 48\,kHz, 20\,mV
amplitude, sinusoidal voltage over the contact, which is in series with a
100\,$\Omega$ resistor. The first and second harmonic of the voltage over the
resistor are measured, from which we obtain the first ($G= \partial I/\partial
V$) and second derivative ($\partial G/\partial V = \partial^{2} I/\partial
V^{2}$) of the current with respect to voltage of the contact. The conductance
is
determined with an accuracy better than 1\% for values larger than $0.5 G_{0}$.
We use a HP3325b function generator to produce the modulation voltage while two
Stanford Research SR830 lock-in amplifiers at $f$ and $2f$, with a time constant
of 10\,ms, are used to obtain the first and second derivative. 16 bit analog to
digital and digital to analog converters are used to control and measure the
piezo voltage. A PC-based controller sweeps the piezo voltage up, and while the
contact breaks, the readings of $G$ and $\partial G/\partial V$ are taken
through an IEEE connection every 100\,ms. A full curve of the contact from a
conductance of over 20\,$G_{0}$ to the transition to vacuum tunneling is
recorded in about 30 seconds.

A large number of such curves have been taken for gold, silver, copper, sodium,
aluminum, niobium and iron. In order to avoid anomalously large $\partial
G/\partial V$ values due to unstable contacts near a conductance step, and to
avoid measuring the average properties of two different plateaus as a result of
the finite integration time of the lock-in amplifiers (which average $G$ and
$\partial G/\partial V$ over 10\,ms), only points on a plateau are included
through exclusion of data points, with suitable selection criteria, for which
the deviation of $G$ and $\partial G/\partial V$ with respect to previous and
consecutively recorded data points is too large. After applying this exclusion procedure to each of these materials, we have analyzed the results by combining all the
data and sorting it according to conductance. Then a fixed number of consecutive
data points were taken from which $\sigma_{GV} = \sqrt{\langle(\partial
G/\partial V)^{2}\rangle-(\langle \partial G/\partial V \rangle^{2})}$ and the
average conductance value were determined. With this method we obtained
$\sigma_{GV}$ as a function of conductance, in a way which is independent of the
number of sampled points at a particular conductance value.

\section*{The noble metals \newline copper, silver and gold}

\begin{figure}[htb]

\epsfig{file=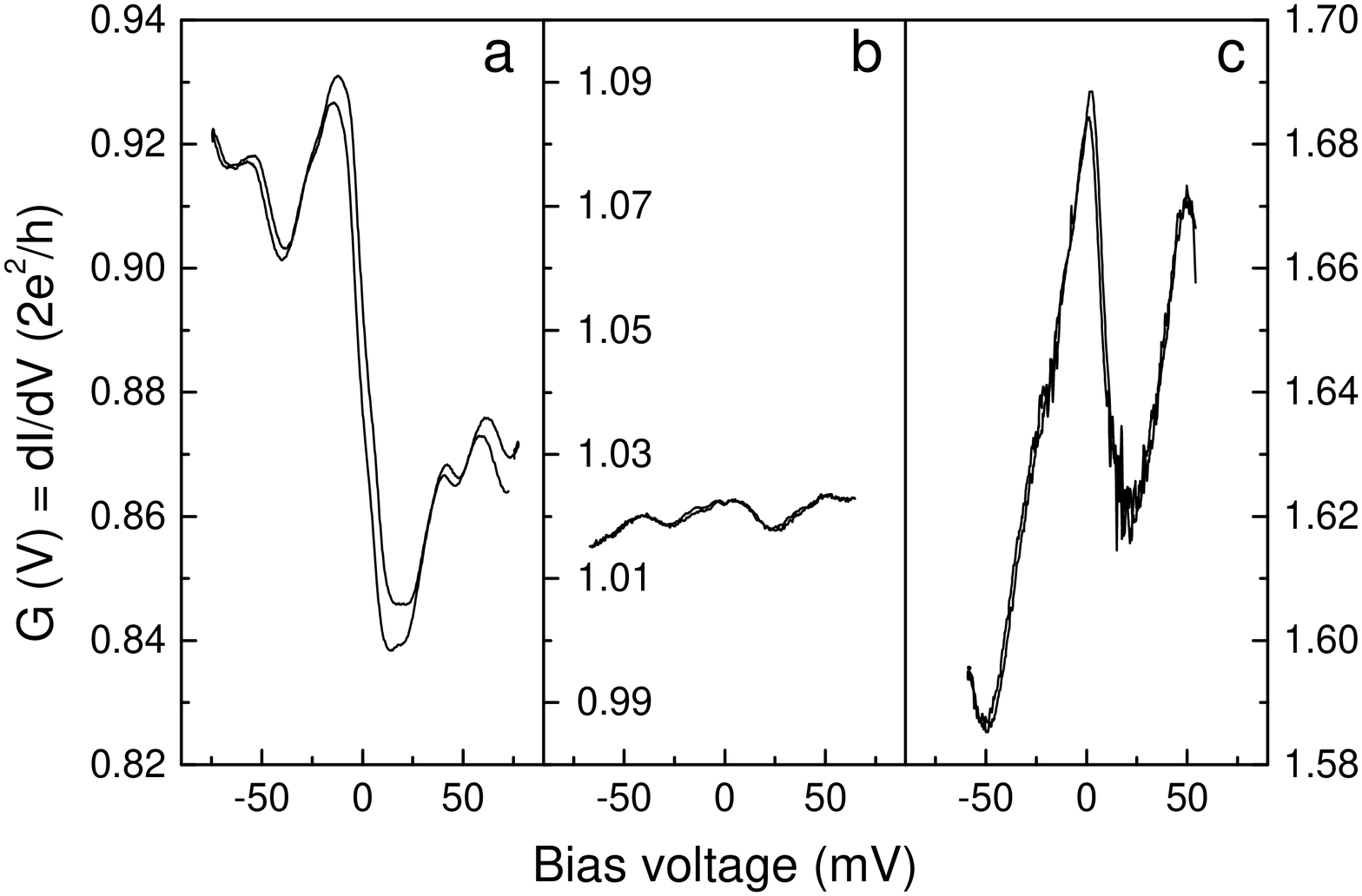,width=5cm,angle=270}

\caption{Plotted in the three panels is the differential
conductance $\partial I/\partial V$ as a function of bias voltage,
measured with a modulation amplitude $<$ 0.35 mV, for three
different gold contacts with $G$ (a) $\sim 0.88 G_{0}$,  (b) $\sim
1.02 G_{0}$, and  (c) $\sim 1.65 G_{0}$. For all three curves the
vertical-scale spans 0.12 $G_{0}$.} \label{fig:fig1}
\end{figure}

The three panels in Fig.\,\ref{fig:fig1} show the measurement of the
differential conductance, obtained with a small ($<$ 0.35 mV) modulation
voltage, against bias voltage for three gold atomic size contacts. In each case
two curves are plotted, one for increasing and one for decreasing bias voltage,
showing the reproducibility of the behavior. The bias
voltage over the contact determines the energy of the electrons injected into
the banks and hence modifies the electron interference resulting from
electrons scattered back towards the contact in the banks. This change in the
interference gives rise to the fluctuations shown in Fig.\,\ref{fig:fig1}. In
our experiments described below the voltage dependence of the conductance is
determined with a
modulation amplitude of 20\,mV, i.e., the average slope of curves such as those
presented in Fig.\,\ref{fig:fig1} is determined over a bias voltage range
of $\pm 20$\,mV. Note the small amplitude of the fluctuations in
Fig.\,\ref{fig:fig1}b. We will argue later that this is an example of the
reduction of $\sigma_{GV}$ for the conductance of gold contacts with value near
1\,$G_{0}$ due to the $\sqrt{T_{n}^{2}(1-T_{n})}$ factor in Eq.\,\ref{sigma}.

\begin{figure} [htb]
\epsfig{file=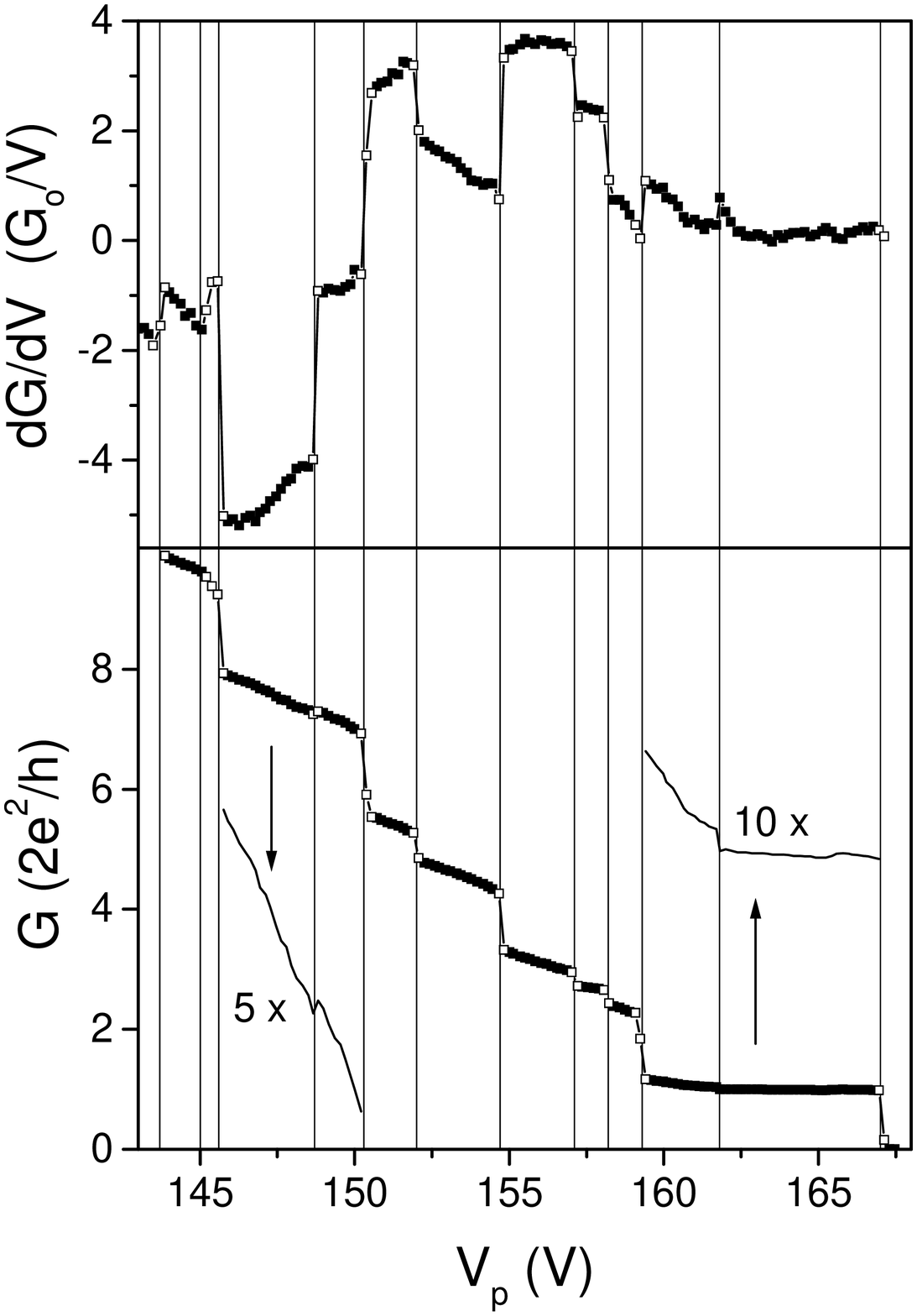,width=7cm,angle=0} \caption{Typical example
of the simultaneous measurement of voltage dependence of the
conductance $\partial G/ \partial V$ and the conductance $G$, as a
function of piezo voltage $V_{P}$ for gold measured with a
constant modulation amplitude of 20\,mV. The graph includes
vertical gray lines which show that the steps in both quantities
coincide. Two plateaus have been enlarged and offset to show the
tiny steps in the conductance. The open squares represent the
points excluded from the ensemble average by the selection
procedure. The elongation of the contact is linear with $V_{P}$
and 10\,V corresponds to about 1\,nm.} \label{fig:fig2}
\end{figure}

An example of the typical conductance and $\partial G/\partial V$
behavior when breaking a gold contact for a constant modulation
amplitude of 20\,mV and zero bias is presented in
Fig.\,\ref{fig:fig2}. The steps and plateaus in the conductance
correspond with atomic rearrangements and elastic deformation
respectively, as the contact is pulled apart and finally breaks
\cite{agrait}. At each step in the conductance we find a
corresponding step in $\partial G/ \partial V$. Even tiny steps in
the conductance, such as between 7 and 8 $G_{0}$, can produce
dramatic jumps in $\partial G/ \partial V$. Changes in electron
path lengths of the order of the Fermi wavelength (which is the
atomic scale for metals) occur at these steps in the conductance
as a result of atomic rearrangements, and randomly change the
resulting electron interference. The continuous change of
$\partial G/\partial V$ during elastic deformation of the contact
along a plateau results from the gradual elongation of the
electron path lengths and hence in a gradual change of the
resulting interference.

In Fig.\,\ref{fig:fig2} the open squares represent the points at
steps in the conductance and $\partial G/\partial V$ which have
been excluded from the statistical analysis by the selection
procedure discussed above. As can be seen in the figure, the
excluded data consists exclusively of the last and first points on
a plateau, together with points which lie between two plateaus as
a result of the finite integration time of the lock-in amplifiers.

\begin{figure} [htb]
\epsfig{file=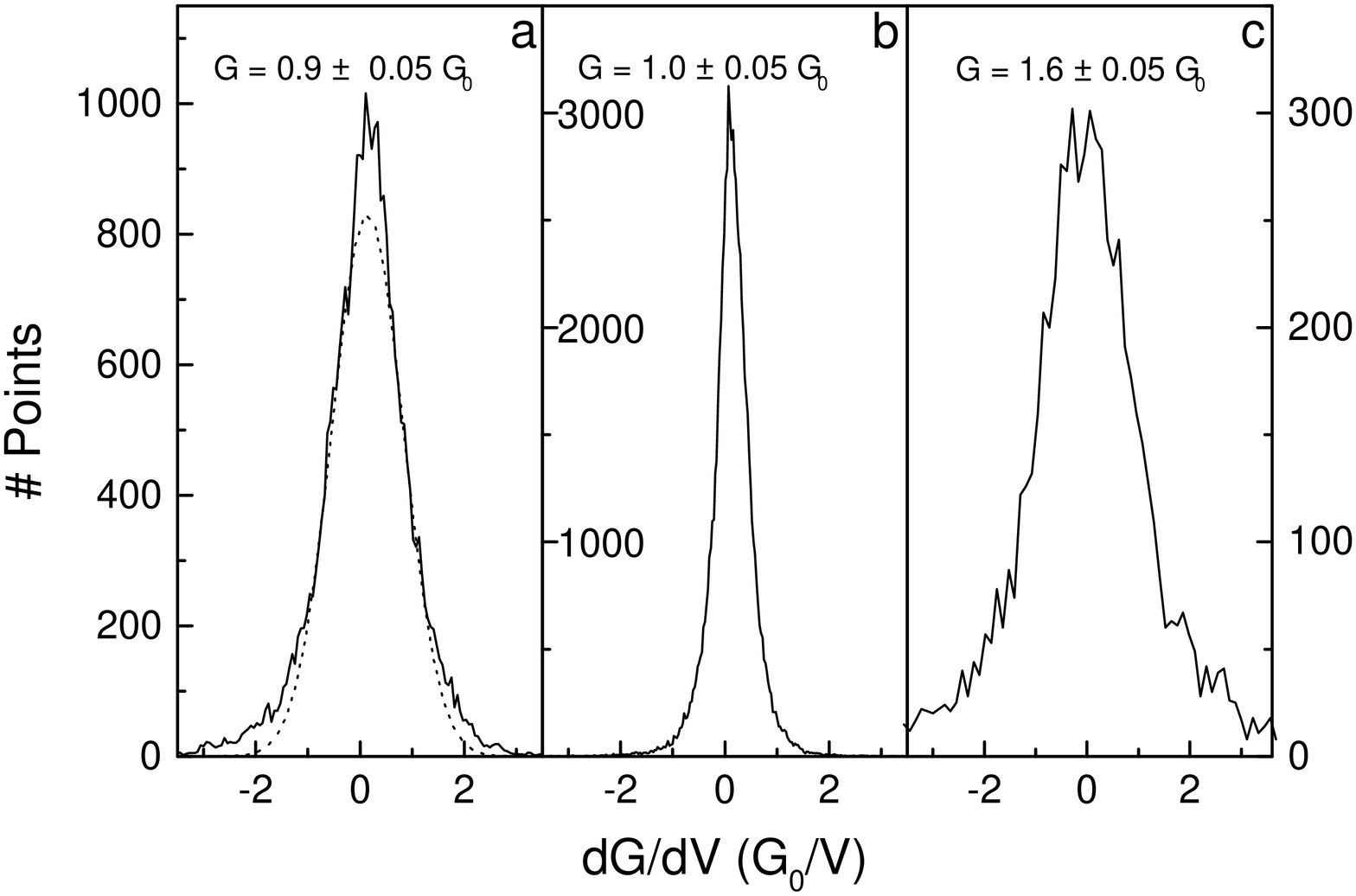,width=5cm,angle=270} \caption{The
distribution of $\partial G/\partial V$ values in a particular
conductance range collected from 3500 individual curves for gold
such as the one presented in Fig.\,3. The conductance range of the
three curves roughly corresponds with the conductance of the
$\partial I/\partial V$ curves in Fig\,2. (a) $G = 0.9 \pm 0.05
G_{0}$ (b) $G = 1.0 \pm 0.05 G_{0}$ (c) $1.6 \pm 0.05 G_{0}$. The
dotted curve in (a) represents a Gaussian fit of the data.}
\label{fig:distrib} \end{figure}

Fig.\,\ref{fig:distrib} shows the distribution of values measured for $\partial
G/\partial V$ in a particular range of conductance collected from 3500
individual curves similar to the one presented in Fig.\,\ref{fig:fig2}. The
distribution at 1\,$G_{0}$ is clearly much narrower than the other two at
non-integer values. This is statistical evidence for what was already observable
in measurements of the bias dependence of the conductance for a single contact
in Fig.\,\ref{fig:fig1}b. We will argue that the narrow distribution can be
explained by a
suppression of the conductance fluctuations (Eq.\,\ref{sigma}), as a result of
$T_{1}$ being approximately equal to 1 and all other $T_{n} \approx 0$. In the presentation of the
data below we concentrate on the width of these types of distributions,
$\sigma_{GV}$, determined for a fixed number of data points.

The shape of the distribution curve shown
in Fig.\,\ref{fig:distrib}a is much sharper around the $\partial G/\partial V =
0$ value than a Gaussian
distribution. The tails of these curves also deviate from Gaussian behavior.
These deviating features are analogous to the peaks calculated and measured in
the distribution of parametric derivatives (e.g. thermopower) of quantum dots
with single mode ballistic point contacts \cite{dots}. The origin of this cusp
at zero amplitude is the limitation of the range over which the differentiated
parameter can vary in value. At both of its maximal values the parametric
derivative is zero, leading to an enhancement of the statistics at zero
amplitude.

\begin{figure} [htb]
\epsfig{file=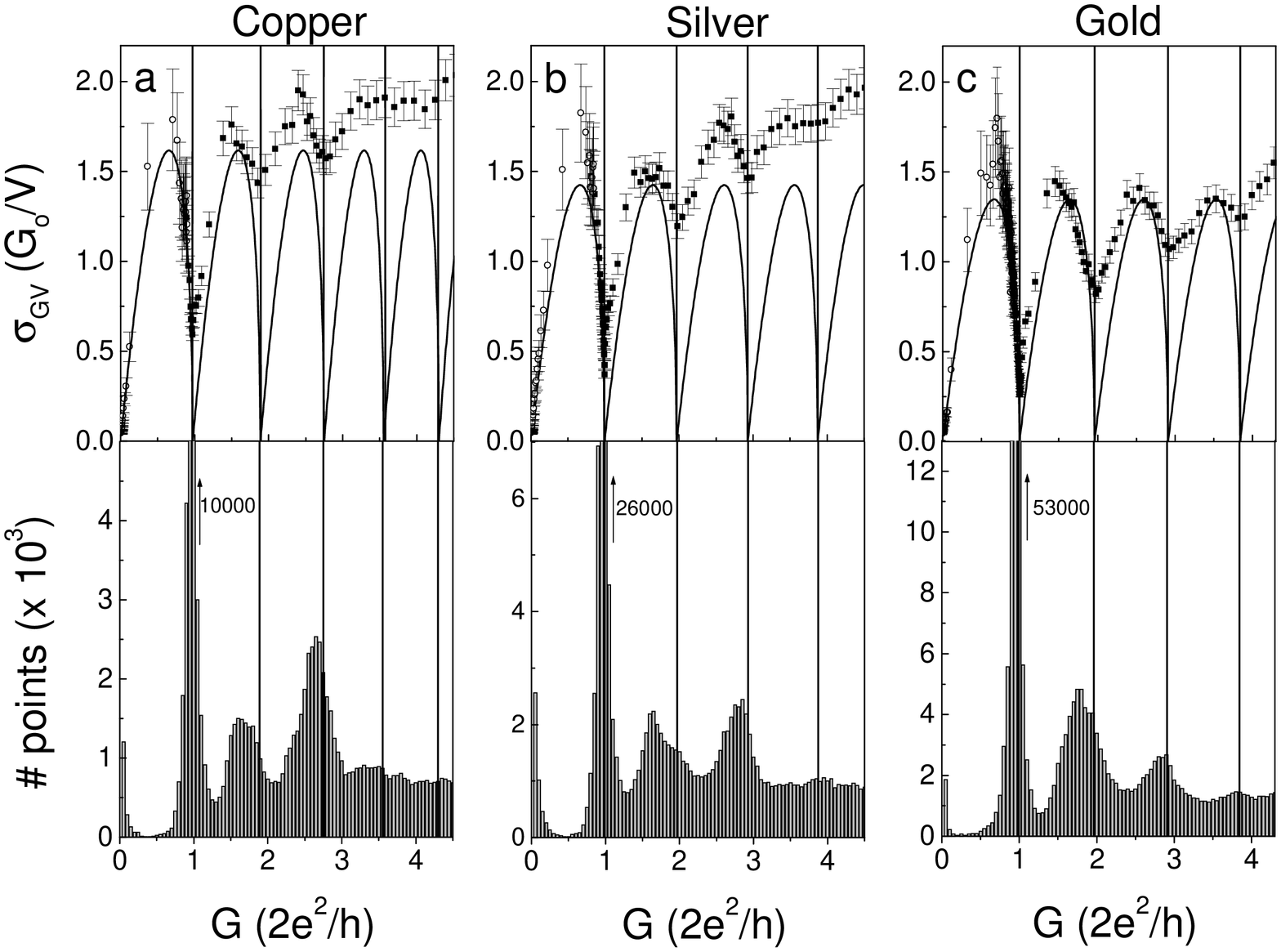,width=6cm,angle=270} \caption{$\sigma_{GV}$
(top) and conductance histogram (bottom) against $G$ for (a)
copper 3000 curves (b) silver 2400 curves (c) gold  3500 curves.
The averages have been obtained from 300 data points for each of
the open circles and 2000, 2000 and 2500 data points for each of
the solid squares for copper, silver and gold respectively. The
solid curve in each case show the behavior of a single channel
opening at a time. The vertical gray lines represent the corrected
integer conductance values.} \label{figAuCuAg} \end{figure}

In the upper panels of Fig.\,\ref{figAuCuAg}, we present the measured
$\sigma_{GV}$ for the noble metals copper, silver
and gold. The data points for $\sim$\,3000 individual curves such as the one in
Fig.\,\ref{fig:fig2} were sorted as a function of conductance. From this total
collection of data points the root mean square of $\partial G/\partial V$ was
calculated for groups of 300, 2000 or 2500 successive data points, depending on
the density of points available. This total collection of data points was also used to calculate the
corresponding conductance histograms plotted in the lower panels of the figure.

The electronic properties of these three noble metals are very similar, which is
reflected in the similar behavior we obtain for $\sigma_{GV}$ as a function of
conductance. Minima in $\sigma_{GV}$ near 1, 2 and 3\,$G_{0}$ can be observed in
all three cases. The minima, however, are most pronounced for gold which even
has a small dip near 4\,$G_{0}$. Another important similarity, as is apparent
from the peaks in the conductance histogram, is the preferred values for the
conductance just below 1, 2 and 3\,$G_{0}$ for all three materials
\cite{krans,brand,gai,costa,sirvent}.

When comparing the experimental results with our model it is important to note
that a given value for $G=G_{0} \sum_{n=1}^{N}T_{n}$ can be constructed in many
ways from a choice of transmission values $\{T_{n}\}$. The experimental values
for $\sigma_{GV}$ are, therefore, an average over impurity configurations {\it
and} transmission values. Assuming the averages are independent, we can compare
the data with various choices for the distribution of the transmissions. In
each upper panel of Fig.\,\ref{figAuCuAg} the solid curve
depicts the behavior of Eq.\,\ref{sigma} for a single channel opening at a time,
i.e. in the interval $G/G_{0} \in \lbrace 0,1 \rbrace$ there is a single channel
contributing to the conductance with $G=G_{0}T_{1}$, in the interval $\lbrace
1,2 \rbrace$ there are two channels with one fully open $G=G_{0}(1+T_{2})$, etc..
The curves in the figure have been scaled so they best fit the data. From this
scaling an estimate for the mean free path can be obtained when a reasonable
value range \cite{agrait2} for the opening angle $\gamma =
35^{\circ} - 50^{\circ}$ is assumed. For copper, silver and gold, we obtain a
value of $l_{e} = 3 \pm 1$\,nm, $l_{e}
= 4 \pm 1$\,nm and $l_{e} = 4 \pm 1$\,nm respectively.

For gold the description of the experimental data with a single
channel opening at a time works surprisingly well In particular,
for the minimum near 1\,$G_{0}$, and for the fact that the maximal
values between the integer conductance values are all nearly
equal. The main discrepancy is that the minima become less
pronounced for higher conductances. The well-developed structure
observed in $\sigma_{GV}$, with a dependence which closely follows
the $\sqrt{\sum T_n^2(1-T_n)}$ behavior of Eq.\,\ref{sigma},
demonstrates a property of the contacts which we refer to as the
saturation of the channel transmission \cite{me}: There is a
strong tendency for the channels contributing to the conductance
of atomic-size contacts of gold to be fully transmitting, with the
exception of one, which then carries the remaining fractional
conductance.

For copper and silver the
amplitude of the data increases together with the degradation of the minima.
These two metals also exhibit the saturation of the channel
transmission effect, but clearly not as rigorous as for gold. This
reflects itself in the estimates we can make for the contribution of an additional
channel at the first three minima. Neglecting the small contribution of the
higher order terms in $a_{l,r_{mn}}$, these are for copper 2\%, 12\% and 15\%,
for silver 1\%, 11\% and 18\% and for gold 0.5\%, 6\% and 10\%.
The concept of
the saturation of transmission channels is consistent with the model of Cuevas
{\it et al.} \cite{cuevas} and other recent experimental work, which shows that,
for gold, the conductance at $G=1$\thinspace $G_{0}$ of a single atom is carried
by a single mode \cite{scheer2,vdBrom}.

The minima in $\sigma_{GV}$ lie at values for $G$ below the integer conductance
values. This shift is due to the scattering of transmitted electrons back to the
contact, which apart from a fluctuating first order contribution in
$a_{l,r_{mn}}$ which determines $\partial G/\partial V$, also gives rise to a shift
in $\langle G \rangle$ when contributions to second order in $a_{l,r_{mn}}$ are
taken into account. Ideally we would like to plot $\sigma_{GV}$ as a function of
$\sum_{n=1}^{N} T_{n}$, with $T_{n}$ the transmission probability of mode $n$ of
the bare contact, but the bare contact is always measured in series with the
diffusive banks. In order to correct for the backscattering to lowest order, the
theoretical curves have been plotted as a function of $G = G_{0}\sum_{n=1}^{N}
T_{n} (1-\sum_{m=1}^{N}T_{m}(\langle |a_{l_{mn}}|^{2}\rangle +
\langle |a_{r_{mn}}|^{2}\rangle))$. The $\langle |a_{l,r_{mn}}|^{2}\rangle $
have been adjusted for optimal agreement with the experimental minima. The
vertical gray lines in each figure represent the corrected integer conductance
values using this lowest order procedure. For gold and silver this value is
comparable $\langle |a_{l,r_{mn}}|^{2}\rangle =0.005$ versus $0.004$
respectively, and hence is consistent with the similar amplitude of $\sigma
_{GV}$ observed for both materials. Copper has a somewhat larger amplitude for
$\sigma _{GV}$, in accordance with a larger shift in the conductance minima with
$\langle |a_{l,r_{mn}}|^{2}\rangle =0.014$. We can
relate this value for the conductance correction to a rough estimate for the
mean free path by equating $\langle |a_{l,r_{mn}}|^{2} \rangle$ to the integral
of
Eq.\,\ref{pcl} over all path lengths. As lower integration limit we have taken
the typical shortest
path time $\tau_{e}$, for the upper limit infinity and for the opening angle we
have assumed the typical range 35$^{\circ} - 50^{\circ} $. The value for
$l_{e}$ we obtain using this method is $4 \pm 1$\,nm, $7 \pm 1$\,nm and $6 \pm
1$\,nm for copper, silver and gold respectively. This is quite close to
the mean free path derived above from the amplitude of
$\sigma_{GV}$ and therefore in accordance with our model. The
values for the mean free path we obtain are much shorter than what is normally
found for bulk samples, and can probably be attributed to surface scattering
near the contact. Assuming surface
scattering is indeed responsible, an important property of the mean free path
which we neglect here is that $l_{e}$ will not
be a constant as a function of the conductance, but rather increase as the
contact diameter becomes larger. This size dependence of the mean free path is
not expected to be very significant in the range of validity of our model, where we
assume a contact diameter $d<L<l_{e}$

The relatively short $l_{e}$ we obtain is responsible for the correction to
the quantized conductances, but it is too long to hold backscatting
responsible for the measurement of the significant frequency with which non-quantized values are measured. Also, if scattering is held primarily responsible
for reducing the conductance from for instance a perfect conductance of
2\,$G_{0}$ to 1.5\,$G_{0}$, then it is not unreasonable to assume that contacts
with a perfect conductance of 1\,$G_{0}$ are reduced to 0.5\,$G_{0}$ with a
probability of the same order of magnitude. This is not observed experimentally
at low temperatures, as contacts with a conductance of 0.5\,$G_{0}$ occur more
than 500 times less frequently for silver and copper than contacts with a
conductance of 1.5\,$G_{0}$ (The formation of atomic chains \cite{chains}
reduces this ratio to about 20 times in the case of gold, since the conductance
of the chains is quite sensitive to distortions making contacts with a
conductance of 0.5\,$G_{0}$ occur with an enhanced frequency).  If, on the other
hand, one assumes that contribution from tunneling, due to for instance
geometrical considerations, are more important, the appearance of non-quantized
values above 1\,$G_{0}$ finds a natural explanation. The formation of geometries
with a conductance smaller than 1\,$G_{0}$ is highly unlikely since the smallest
contact geometry is that of a single atom with conductance 1\,$G_{0}$ and when
the contact breaks, the banks relax back preventing high transmission
probability tunneling contributions from contributing. The latter process is
usually referred to as the jump to tunneling \cite{krans1}.

An important feature for all three noble metals is that the minima in
$\sigma_{GV}$ near 2\,$G_{0}$ do not coincide with the respective peaks in the
histograms. The minima lie at the expected conductances based on the
backscattering amplitude we require to consistently fit all the minima, the so
called corrected integer conductance values. The second peak in the histograms
clearly are located at lower conductance values. We propose that this
discrepancy
is caused by favorable atomic configurations which have a bare conductance
smaller than 2\,$G_{0}$, and thus give rise to a peak in the histogram below the
corrected conductance value for 2\,$G_{0}$.

\section*{Test of the model}

Apart from the set of transmission values $\lbrace T_{n} \rbrace$ the equation
contains two free parameters, $l_{e}$ and $\gamma$, over which we do not have
any experimental control. The dependence on the modulation voltage $V_{mod}$, however, is a
parameter which we can verify, assuming that all other relevant parameters are
independent of the applied voltage. This is a reasonable assumption for the
modulation amplitudes at which we measure, as parameters like $\lbrace
T_{n} \rbrace$ vary on the scale of $E_{F}$, and $l_{e}$ and $\gamma$ are not
expected to change on the scale of 100mV. Assuming further that $eV_{mod} >
\hbar/\tau_{\phi}$
then the product $\sigma_{GV} V_{mod}^{3/4}$ should be constant for all
$V_{mod}$.

\begin{figure} [htb]
\epsfig{file=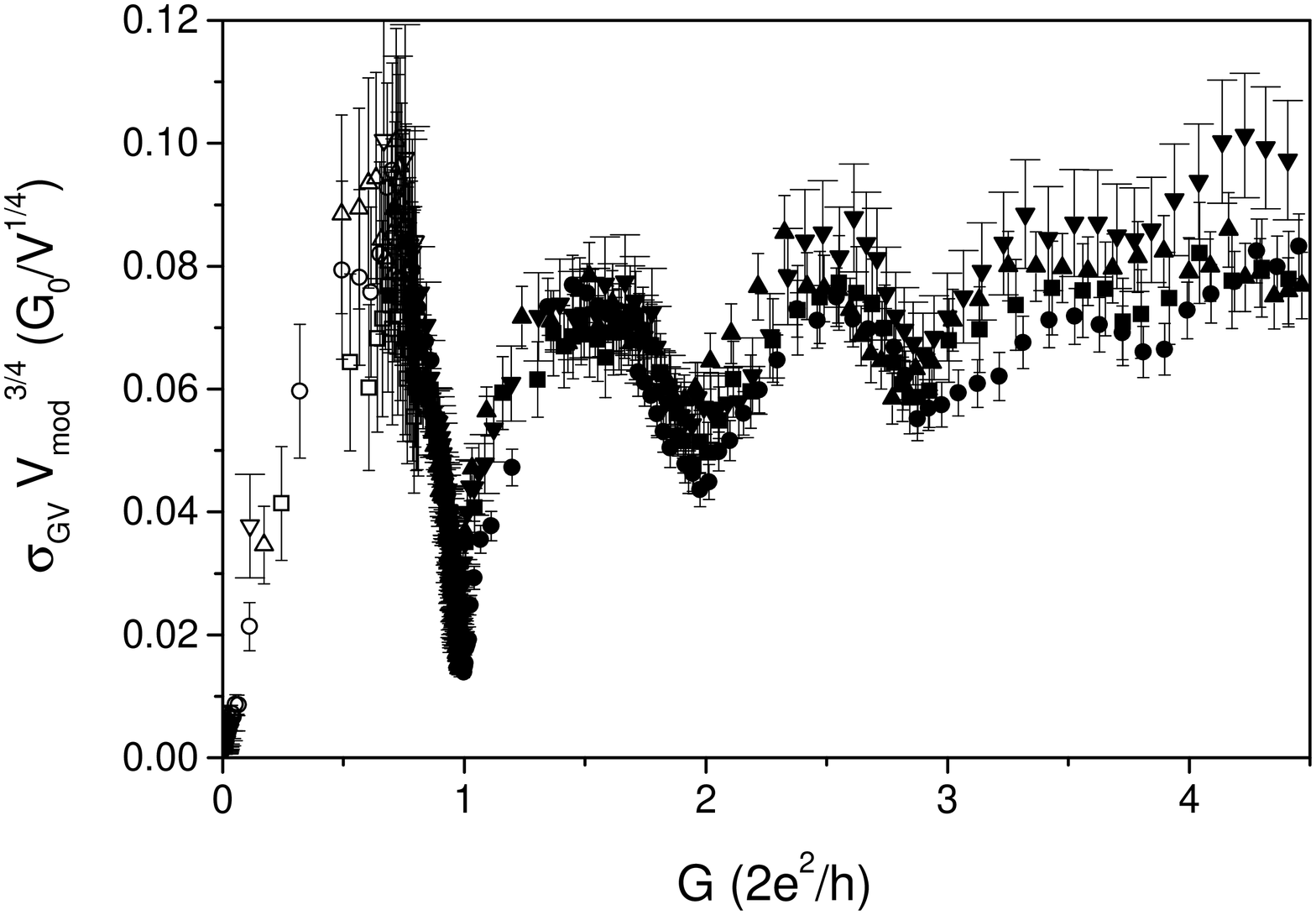,width=5cm,angle=270} \caption{$\sigma_{GV}$
multiplied by the modulation amplitude to the power $3/4$ against
conductance $G$ for gold measured at (squares) 10\,mV, (circles)
20\,mV, (up triangles) 40\,mV, (down triangles) 80\,mV.}
\label{figV} \end{figure}

In
Fig.\,\ref{figV}, $\sigma_{GV} V_{mod}^{3/4}$ has been plotted against $G$, for
$V_{mod}$\,=\,10, 20, 40 and 80\,mV. Within the experimental
accuracy no modulation voltage dependence is observed, as all four data sets coincide very well. This would not be
the case unless the power of the modulation amplitude dependence is close to 3/4. Using a procedure that calculates the minimal difference between
the six combinations of experimental curves, which have been multiplied by their
respective modulation amplitudes to a power which is the free parameter, we find
this power to be $0.71 \pm 0.06$ in good agreement with the $3/4$ predicted by
the theory.

The mechanism used to describe the fluctuations in the conductance above also
produces fluctuations in other transport properties, notably the thermopower.
Measuring the thermopower of atomic size metallic contacts requires a completely
different experimental method and is performed on an energy scale much smaller
than that necessary for determining $\sigma_{GV}$. The experimental results \cite{thermo},
however, have been successfully described by a theory based on the same
principles as those presented above. The predicted theoretical
relationship between the standard deviation of the thermopower $\sigma_{S}$ and
$\sigma_{GV}$ is given by,
\begin{equation}
\sigma _{GV}=\sigma _{S}\frac{2.71e^{2}G_{0}\sum_{n=1}^{N}T_{n}}{
ck_{B}(k_{B}\theta )^{1/4}(eV_{mod})^{3/4}},  \label{eqsigssigg}
\end{equation}
where the numerical constant $c=5.94$. The only parameters in the above relation are, $\theta$, the temperature at
which the thermopower measurements were performed, $V_{mod}$, the modulation
amplitude for the conductance fluctuation measurements and the total conductance
$G_{0}\sum_{n=1}^{N}T_{n}$. All unknown
parameters in Eq.\,\ref{sigma}, the set of transmission probabilities $\lbrace
T_{n} \rbrace$, $l_{e}$ and $\gamma$ cancel out, and the scaling relation
provides an independent test of the experimental data.

\begin{figure} [tb]
\epsfig{file=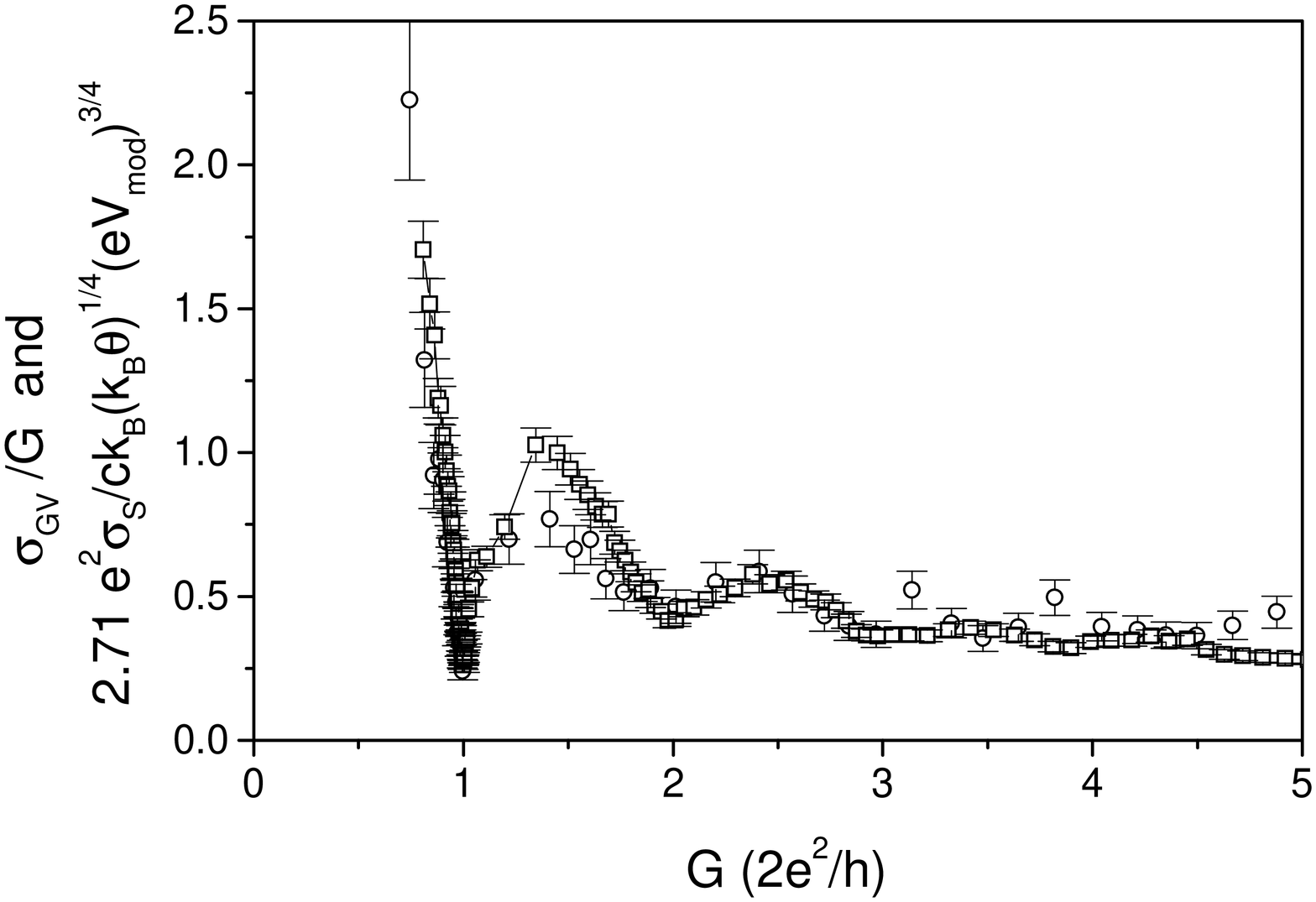,width=5cm,angle=270} \caption{Comparison of
the standard deviation of the thermopower $\sigma _{S}$ and the
standard deviation of the voltage dependence of the conductance
$\sigma _{GV}$ by plotting ($\bigcirc$) $2.71e^{2}\sigma
_{S}/ck_{B}(k_{B}\theta )^{1/4}(eV_{mod})^{3/4}$ and ($\Box$)
$\sigma _{GV}/G$ as a function of conductance.} \label{figsvsg}
\end{figure}
By comparing $\sigma_{GV}$
to measurements of the thermopower on atomic size contacts \cite{thermo} using
Eq.\,\ref{eqsigssigg} we can effectively extend the energy range over which we
test our model to an order of
magnitude smaller, and test the experimental procedure against a completely
independent method. In Fig.\,\ref{figsvsg} we have plotted $2.71e^{2}\sigma
_{S}/ck_{B}(k_{B}\theta )^{1/4}(eV_{mod})^{3/4}$ and $\sigma _{GV}/G$ as a
function of conductance. We have set $\theta = 12$\,K
and $V_{mod} = 20$\,mV in accordance with the experimental conditions.
Excellent agreement is obtained between both
experimental methods {\it without any free parameters}. We interpret this as a
successful test for the validity of the principle on which our theoretical
analysis is based. The dependence on the opening angle $\gamma$, however we
regretfully cannot verify experimentally as we have no control over the contact
geometry.

\section*{The alkali metal sodium}

\begin{figure} [tb]
\epsfig{file=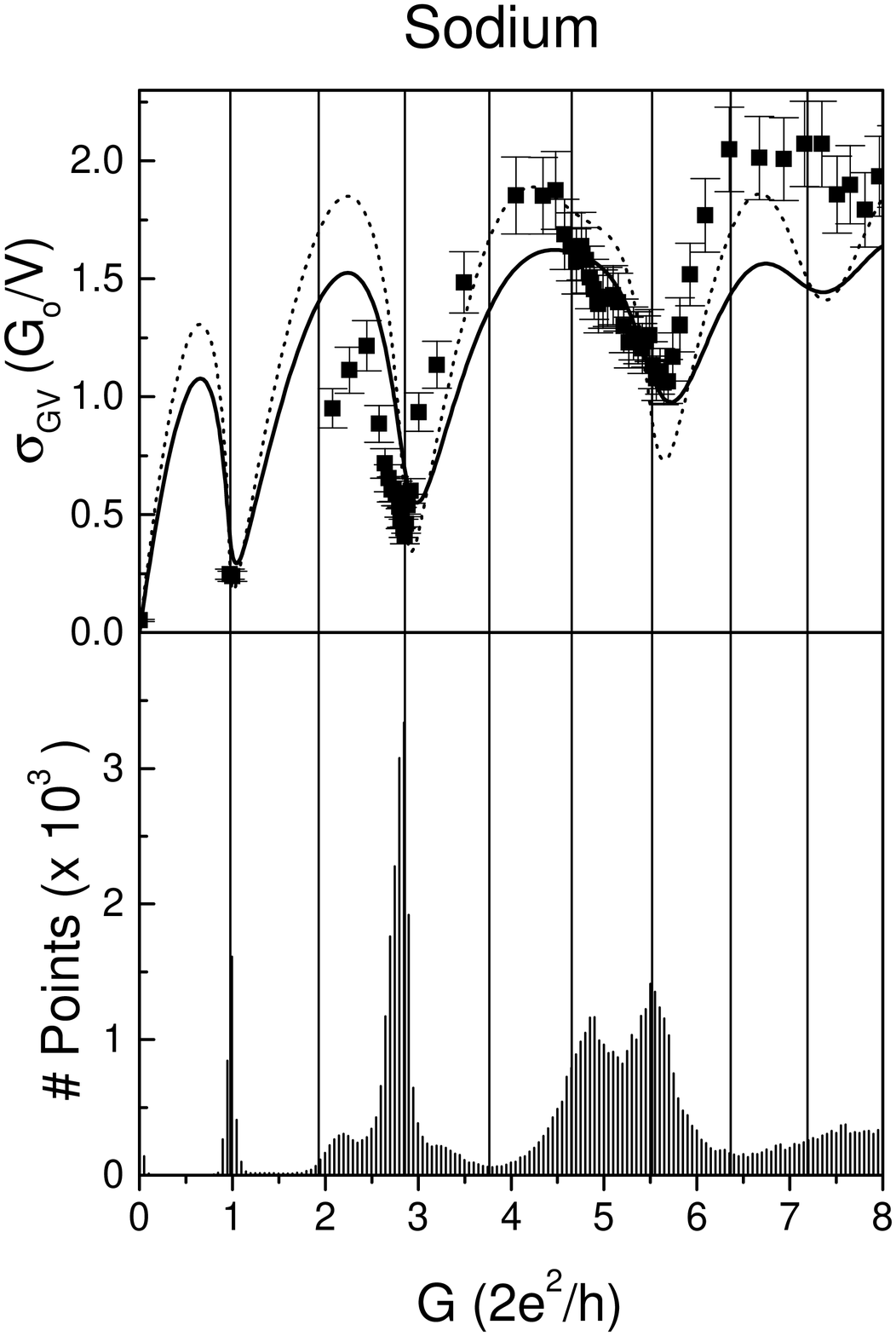,width=7cm,angle=0} \caption{$\sigma_{GV}$
(top) and conductance histogram (bottom) against $G$ for 1800
sodium curves with each solid square representing the statistics
on 1000 data points. The vertical black and gray lines indicate
the corrected integer conductance values for which the histogram
peaks respectively do and do not correspond with minima in
$\sigma_{GV}$. The curves depict the behavior of a hyperbolic
constriction in a three dimensional electron gas with circular
aperture, with (---) opening angle $\gamma = 60^{\circ}$ and mean
free path $l_{e} = 4.4$\,nm, ($\cdot$\,$\cdot$\,$\cdot$) $\gamma =
45^{\circ}$ and $l_{e} = 5$\,nm.} \label{figNa} \end{figure}

Sodium also is a monovalent metal but its histogram determined from 1800 curves
(lower panel Fig.\,\ref{figNa}), is completely different from that observed for
copper, silver or gold. The
statistically preferred conductance values are observed as peaks in the
histogram \cite{krans} near 1, 3, 5 and 6\,$G_{0}$ rather than near 1, 2,
3\,$G_{0}$. This
series of peaks in the histogram at 1, 3, 5 and 6\,$G_{0}$ have been interpreted
as resulting from the quantization of the conductance in a cylindrically shaped
nanowire. The histogram peaks are very sharp, and in the 1800 curves measured
almost no data is obtained between 0 and 1\,$G_{0}$ and between 1 and
2\,$G_{0}$. For this reason no points for $\sigma_{GV}$, determined from the
same 1800 curves, are presented in these ranges (upper panel Fig.\,\ref{figNa}).
Even with these points absent, the $\sigma _{GV}$ measured for sodium is
distinctly different from that observed for the noble metals. In $\sigma_{GV}$
we observe definite minima near 3 and 6\,$G_{0}$ and
although there has been no data measured in the ranges $0 < G < 1$\,$G_{0}$ and
$1
< G < 2 G_{0}$, the value of $\sigma_{GV}$ at 1\,$G_{0}$ is small making it
a very
probable location of a minimum. Since there is no data below
2\,$G_{0}$, we cannot exclude that there is a small minimum at 2\,$G_{0}$.

The histogram peaks coincide with the minima in $\sigma_{GV}$,
with the exception of the peak near 5\,$G_{0}$. The absence of a
minimum at 5\,$G_{0}$ is at first surprising. When one considers
that in a conductance histogram for a model of a three dimensional
cylindrical contact based purely on a free electron gas, the peak
at this value is found to be nearly absent due to smearing by
tunneling contributions \cite{torres}, it is striking that a
histogram peak is there at all. Also, unlike the other peaks, the
one below 5\,$G_{0}$ and also the small one above 2\,$G_{0}$ do
not coincide with the corrected integer conductance values. We
propose that these two peaks result from favorable atomic
configurations, which are sampled more often than other
conductance values while making a histogram, but do not result
from stable quantized conductance values determined by an integer
number of nearly open channels.

As in the case of the noble metals there
is a systematic shift of the position of the minima in $\sigma_{GV}$ to lower
conductance values. The corrected integer conductance multiples are shown in
Fig.\,\ref{figNa} as vertical gray and black lines depending on whether they
coincide with the minima in $\sigma_{GV}$ or not. For the correction to the
integer conductance values in Fig.\,\ref{figNa} we have used Eq.\,\ref{rcorr}
with $r = 0.015$. This value corresponds to a series
resistance of
200\,$\Omega$ for $g \to \infty$ and 400\,$\Omega$ for $g = 1$ as is evident from Eq.\,\ref{rcorrsimp}.
Eq.\,\ref{rcorr} is used, rather than the first order correction applied to the
noble metals, because the conductance extends to larger values where higher
order terms in the backscattering amplitude become relevant.

We want to compare these results with the simplest possible model, which
neglects the atomic character of the contact and only takes the
cylindrical
symmetry and finite length of the contact into account. For this purpose we have made use of the
calculations by Torres {\it et al.} \cite{torres2}. The model consists of a free
electron gas confined by hard wall boundaries, which have the form of a
hyperboloid. The differential equations for this system can be solved
numerically, from which the transmission probabilities $T_{n}$ for each mode as
a function of the contact diameter can be obtained. This makes a direct
evaluation of Eq.\,\ref{sigma} possible, where the only remaining adjustable
parameter is the mean free path. The opening angle which describes the shape of
the hyperboloid and thus the dependence of the mode transmissions on the contact
diameter, is the same as the one which enters in Eq.\,\ref{sigma}. We have added
two $\sigma_{GV}$ curves calculated for such a system to the graph. For curves
in the opening angle range from 60$^{\circ}$, with mean free path
$l_{e}=4.4$\,nm, to 45$^{\circ}$, with $l_{e}=5$\,nm, we find reasonable
agreement between various ranges in the data and the theoretical curve.

The differences between the calculated curve and the measured data can be attributed to the averaging over many contact geometries and thus over a range of $\gamma $ values. Also, the smearing in the conductance $\langle
G \rangle$ due to
the ensemble average of defect configurations is not included. This
property will make the minima less deep and sharp but will hardly influence the
maxima.

Another feature of the calculated curve, which can also be recognized in the

measured data, is that the minimum in the experimental and calculated $\sigma_{GV}$ below 6\,$G_{0}$ does not coincide exactly with the corrected quantized value for the conductance, even after
application of the same series resistance we have used to compensate the shift
in the histogram peaks. In other words, when we ignore the series resistance correction, the model predicts the minima to be shifted above the integer values. This is a direct result of the significant tunneling
contributions for the opening angles we have used to
model the contact when just opening a mode in combination with the asymmetry of
the dependence of
$\sigma_{GV}$ on the mode transmission. This systematic shift becomes more
pronounced with larger opening angles, larger conductance values and the
presence of degenerate modes. The value of the opening angle we obtain for
sodium from the theoretical curves, $\gamma = 45^{\circ} - 60^{\circ}$ is
comparable but somewhat larger than the typical estimates made for the opening
angles of atomic size
gold contacts \cite{agrait2}.

\section*{The trivalent metal aluminum}

\begin{figure} [htb]
\epsfig{file=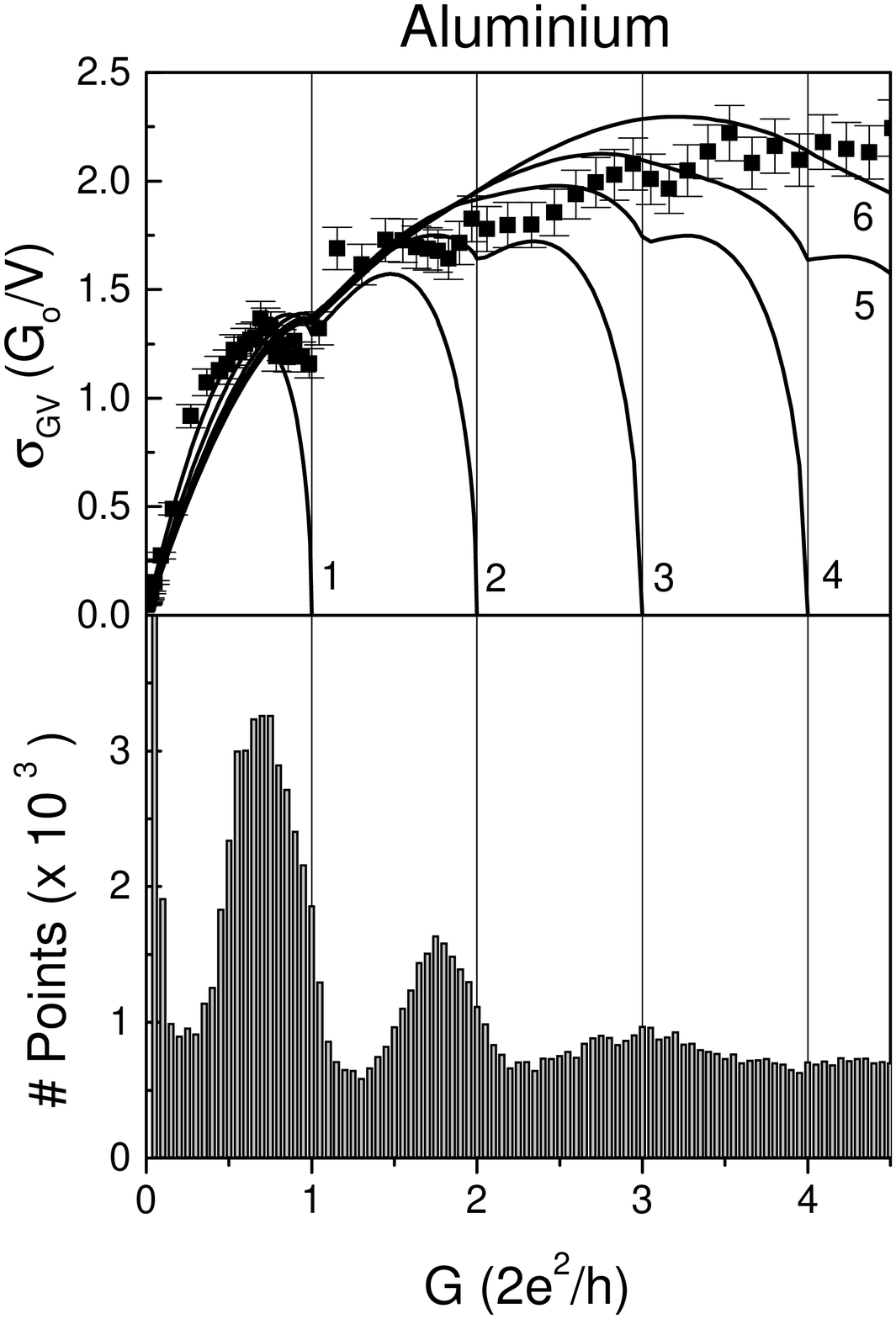,width=7cm,angle=0} \caption{$\sigma_{GV}$
(top) and conductance histogram (bottom) against $G$ for 2800
aluminum curves with each solid square representing the statistics
on 2000 data points. The curves in the graph have been labeled
with a number and represent the contribution to $\sigma_{GV}$ of
(1) a single channel opening, a random distribution over (2) two,
(3) three, (4) four, (5) five and (6) six channels (see text).}
\label{figAl} \end{figure}

The statistically preferred conductance values for aluminum are shown in the
lower panel of Fig.\,\ref{figAl}. The clear peaks, evident in the histogram below
1 and 2\,$G_{0}$ and a weak bump above 3\,$G_{0}$ are in accordance
with previous measurements \cite{yanson}. The peaks are less pronounced, but at
first glance similar to those observed for gold, silver and copper. The most
important discrepancy between the monovalent metals and aluminum is that in the
latter case the first peak is broader and clearly displaced below 1\,$G_{0}$.

The measured $\sigma _{GV}$ for aluminum, presented in the upper panel of
Fig.\,\ref{figAl}, is completely different from the behavior observed for the
noble metals copper, silver and gold. The clear minima at 1, 2 and 3\,$G_{0}$
have been
replaced by a slight dip at 1\,$G_{0}$. In order to understand these measured
features it is important to realize that a single aluminum atom has a
conductance close to $G_{0}$ but admits three conductance channels
\cite{scheer2,scheer}. It is thus not surprising that the behavior associated
with the
saturation of a single partially open channel is not observed for aluminum. The
histogram peaks observed for this trivalent material can thus be attributed to
another mechanism. A likely candidate is favorable atomic configurations, which
are probed more frequently than others.

 In Fig.\,\ref{figAl} a series of curves
are included which show the behavior for a single channel, and a
random distribution of two, three, four,  five and six channels.
The curves have been generated by calculating the square root of
$\sigma_{max}^{2}\int_{0}^{1}..\int_{0}^{1} P(T_{1},T_{2}, ..,
T_{N})\sum_{n=1}^{N} T_{n}^{2}(1-T_{n})dT_{1}..dT_{N}$ where $P$
is the probability distribution giving an equal probability to
every transmission value for each $T_{n}$ under the constraint
that $\sum_{n=1}^{N} T_{n} = G/G_{0}$. These curves have been
labeled 1, 2, 3, 4, 5, and 6 respectively, and have all been
scaled with the same amplitude. The dips in the calculated curves
with a random distribution of two or more channels results from
the property that, in a random distribution, at multiples of the
conductance quantum, there is a finite probability to encounter
some $T_{n}$ = (0 or 1), for which their contribution to
$\sigma_{GV} = 0$. This effect however, and hence the dip, becomes
less pronounced with an increasing number of channels. For the
lowest conductances ($< 0.5$\,$G_{0}$) one can observe that the
behavior most closely follows that of a single channel. This is
expected as for small conductances in the tunneling regime a
single channel is expected to dominate the conductance and, hence,
the behavior in $\sigma_{GV}$. As the conductance increases the
behavior becomes more like a random distribution over an
increasing number of channels. The random distribution of channels
which we have introduced serves only to illustrate that the small
dip at 1\,$G_{0}$ appears for a limited number of channels even
without the saturation of channel transmission effect, and that
the gradual increase in $\sigma_{GV}$ is a direct result of more
channels contributing for larger conductances. The actual behavior
of the transmission channels is probably not completely random as
can be judged from the theory \cite{cuevas} which shows that there
is usually one dominant channel and two smaller ones for a single
aluminum atom. Nonetheless the curves reproduce the evolution of
$\sigma_{GV}$ as a function of conductance with reasonable
accuracy.

\section*{The transition metals \newline niobium and iron}

\begin{figure} [htb]
\epsfig{file=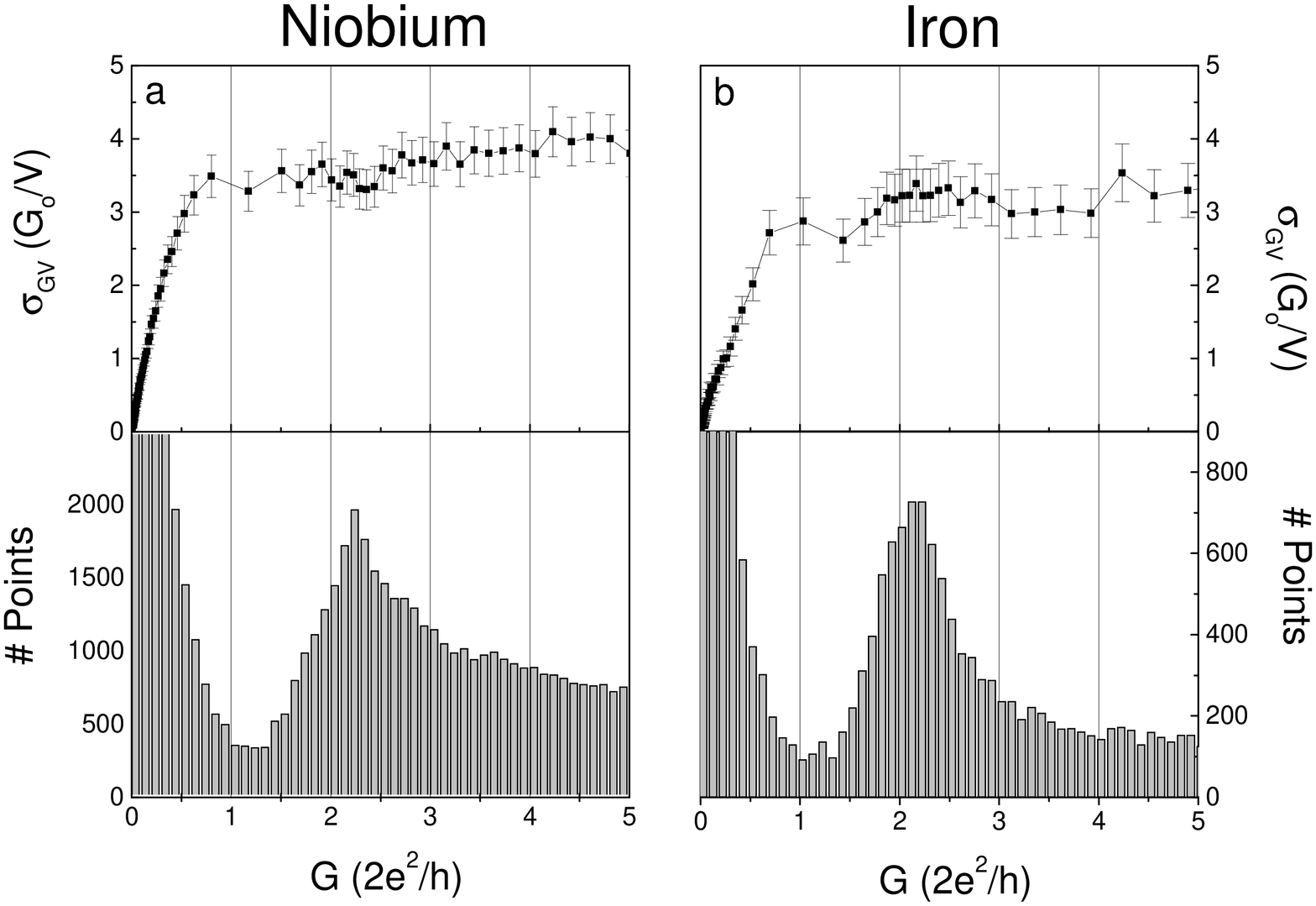,width=5.5cm,angle=270} \caption{(a)
$\sigma_{GV}$ (top) and conductance histogram (bottom) against $G$
for 2400 niobium curves measured at 10\,K with each solid square
representing the statistics on 1000 data points. (b) $\sigma_{GV}$
(top) and conductance histogram (bottom) against $G$ for 700 iron
curves with each solid square representing the statistics on 500
data points.} \label{figNb} \end{figure}

The measured $\sigma _{GV}$ (upper panel) and histogram (lower panel) for 2400
niobium curves, recorded at a temperature of 10\,K in order to avoid effects of
the superconductivity on the voltage dependence of the conductance, have been
plotted in Fig.\,\ref{figNb}a. For iron the measurements of $\sigma _{GV}$
(upper panel) and the histogram (lower panel) recorded for 700 curves are
presented in Fig.\,\ref{figNb}b.

Both $sd$-metals, niobium and iron, show completely different features when
compared to the other materials
we have discussed so far. When compared with each other, the measurements for
niobium and
iron are so similar that they nearly are indistinguishable. For both metals,
$\sigma_{GV}$ increases strongly from 0 to 1\,$G_{0}$ and above this conductance
value, increases only slightly. The dip observed in $\sigma_{GV}$ for
aluminum is absent and the increase in
$\sigma_{GV}$ with conductance is much smaller than was the case with aluminum.
From the single peak in the histogram we can deduce that both materials have a
statistically preferred conductance value just above 2\,$G_{0}$. This peak is
expected to be the result of the reproducibility in the conductance of the last
plateau consisting of a single niobium or iron atom. In the case of niobium this
value is in excellent agreement with the measured and calculated conductance
value for a single atom of niobium (2 to 3\,$G_{0}$ \cite{cuevas,scheer2}). For
iron these calculations and measurements have not yet been performed.

The completely random distribution used to describe the behavior of aluminum
clearly cannot reproduce the measured $\sigma _{GV}$ for niobium and iron. If
one considers distributions for the transmission probabilities that are closer
to the five calculated transmission probabilities contributing to a single-atom
contact \cite{cuevas}, the experimental behavior up to 3\,$G_{0}$ can in
principle be simulated. The number of free parameters in such an analysis,
however, makes such an ad hoc procedure quite meaningless. More theoretical work
must be performed to provide an approximation for the range of transmission
channel distributions that should be considered in order for the measurements to
be reliably related to a general trend. The fact that the measured $\sigma_{GV}$ above 1\,$G_{0}$ is almost not dependent on the total conductance suggests that the number of partially open channels contributing to the conductance does not depend on the contact size.

The histogram we have recorded for
iron is different from the histogram constructed from 80 iron
contacts recorded under ambient conditions in the magnetically saturated state
\cite{fe}. Possibly the different temperature at which the experiments are
performed or the magnetization state are responsible for the discrepancy. The
influence of these experimental conditions should be studied in more detail in
the future.

\section*{Conclusions}

With the technique of ensemble averaged conductance fluctuations we are able to
measure the average properties of the conductance mode evolution of atomic size
contacts. We have successfully tested the modulation voltage dependence of the
theory, and can relate $\sigma_{GV}$ to the standard deviation in the thermopower
without any free parameters. An important property of the measurement of the
conductance fluctuations is that it is not dependent on preferential contact
configurations.

Sodium, the most free electron like material studied, exhibits electronic
behavior which can be reproduced reasonably well with a hyperbolic constriction
in a three dimensional electron gas with a circular orifice. The conductance
histogram of this material however, contains contributions from such a circular
constriction in a three dimensional electron gas as well as other peaks possibly
resulting from favorable atomic
configurations. The conductance properties of the other monovalent materials
such as gold, silver and copper seem to be best described by the tendency of the
conductance channels to open one by one, a property which has been called the
saturation of channel transmission \cite{me}. The conductance histogram of these
materials contains features which seem to coincide with the evolution of the
conductance modes, but particularly the second peak in the histograms are also
determined by other statistical (probably atomic) properties of the contact. For
aluminum, niobium and iron we find the behavior for $\sigma_{GV}$ we expect
based on
the atomic orbital model \cite{cuevas,scheer2,scheer}.
The conductance histograms of these three materials seem to be dominated by the
statistical distribution of atomic configurations.

The concept of saturation of channel transmission has been introduced in
Ref.\,\onlinecite{me}, to make a marked distinction between the properties of
the conductance modes we observe here, and the statistically preferred
conductance values which in the literature are generally referred to as
conductance quantization. Indeed, both are based on the quantum mechanical
Landauer-B\"{u}ttiker formalism, but the latter assumes there is a statistical
preference for contacts with quantized values as a result of this formalism.
Since it is not clear to what extent favorable atomic configurations are
responsible for the histogram peaks, we feel a sharp distinction should be made
between results that {\it can} be influenced by favorable atomic positions {\it
possibly} mimicking the features of conductance quantization, and results which
truly probe the electronic properties of the contact. With the saturation of the
channel transmission we wish only to describe the evolution of these modes, but
we do not rule out that conductance quantization may prove to be an
important factor which influences the contact formation \cite{doscalc}. However, in view of the results presented in Figs.\,\ref{figAl} and \ref{figNb} together
with the arguments presented above for the other materials, we should be aware
that peaks in a histogram by
themselves give no unambiguous information about the actual composition of the
conductance modes.

For the correction to the bare contact conductance due to scattering near the
contact, we find that this correction is correlated to the amplitude of the
conductance fluctuations and hence the elastic mean free path. This provides
strong experimental evidence that these types of scattering effects are indeed
responsible for the so-called series resistance.

\acknowledgments

\noindent This work is part of the research program of the ''\nobreak{Stichting}
FOM'', which is financially supported by NWO. The development of the theory and
interpretation of the gold data was done in collaboration with C. Urbina, D.
Esteve and M. Devoret. We acknowledge the stimulating support of L.J. de Jongh,

\section*{Appendix}

Incorporating a finite modulation voltage into the expression for $\sigma_{GV}$
is somewhat more technical. In analogy with Eq.\,\ref{deltag} we can write for the
energy dependent part of the current,

\begin{eqnarray}
\delta I &=&\frac{2e}{h}\int_{0}^{eV}\sum_{n=1}^{N} T_{n}
2\text{Re}[r_{n}a_{l_{nn}}(-E) \nonumber \\
& & +r^{\prime}_{n}a_{r_{nn}}(eV-E)]dE. \label{apdi}
\end{eqnarray}

\noindent We must include the modulation voltage $V = V_{0} +
V_{mod}\text{sin}(\omega_{0}\tau^{\prime})$ in the argument for the backscattering amplitude,
which we write as an integral over contributions of path traversal times $\tau$.
Hence $ a_{l,r_{nn}}(E)  = \int a_{l,r_{nn}}(\tau) \text{exp}[-iE\tau/\hbar]
d\tau$ where we assume the dominant energy dependence is the phase factor. We
will first consider the contribution to the
current of a single path with time $\tau$: $\delta I(\tau)$. At a later stage we will perform the integration over
different paths. Evaluating the integral over $E$ in Eq.\,\ref{apdi} gives,

\begin{eqnarray}
 \delta I (\tau)&=&const.+ \frac{2e}{h}2 \sum_{n=1}^{N}T_{n}\frac{\hbar }{\tau }
\label{deltaitau}
\\
& &\times \text{Re}[-ir_{n}a_{l_{nn}}(\tau)\text{e}^{ieV_{0}
\tau /\hbar }\text{e}^{+ieV_{mod}\tau \text{sin} (\omega_{0}\tau^{\prime}
)/\hbar }
\nonumber \\
& &+ ir^{\prime}_{n}a_{r_{nn}}(\tau)\text{e}^{-ieV_{0}\tau /\hbar }\text{e}^{-
ieV_{mod}\tau \text{sin} (\omega _{0}\tau^{\prime} )/\hbar }] . \nonumber
\end{eqnarray}

\noindent We are measuring the {\it ac} component of the current at twice the
modulation frequency, this is equivalent to measuring the second derivative of
the current with respect to voltage. Therefore we expand the exponential term
Eq.\,\ref{deltaitau} in harmonics of the modulation frequency,
\begin{eqnarray}
& & e^{ieV_{mod}\tau\text{sin}(\omega_{0}\tau^{\prime})/\hbar}= \sum_{n=-
\infty}^{\infty }e^{in\omega_{0} \tau^{\prime} }J_{n}(eV_{mod}\tau/\hbar)
\nonumber \\
&&= J_{0}(eV_{mod}\tau/\hbar)+
2i\text{sin}(\omega_{0}\tau^{\prime})J_{1}(eV_{mod}\tau/\hbar) \nonumber \\
& & + 2\text{cos}(2\omega_{0} \tau^{\prime})J_{2}(eV_{mod}\tau/\hbar)+...
,\label{bessel}
\end{eqnarray}

\noindent where $J_{n}(z)$ is the $n$'th Bessel function. In the last step we
used $J_{-n}(-z) = (-1)^{n}J_{n}(z)$. We are particularly interested in the last
term in Eq.\,\ref{bessel}, which we will use to obtain the part of the current
which is proportional to twice the modulation frequency ($\text{cos}(2\omega
_{0}\tau^{\prime})$),

\begin{eqnarray}
&&\delta I_{2\omega_{0}}(\tau) =-\frac{2e}{h}4\sum_{n=1}^{N}
T_{n}\frac{\hbar}{\tau }\text{cos}(2\omega_{0}\tau^{\prime})J_{2}(eV_{mod}\tau
/\hbar)  \nonumber \\
& & \times \text{Re}[-ir_{n}a_{l_{nn}}(\tau)e^{ieV_{0}\tau /\hbar
}+ir^{\prime}_{n}a_{r_{nn}}(\tau)e^{-ieV_{0}\tau /\hbar }].
\label{twoomega}
\end{eqnarray}

\noindent In general, the Taylor expansion of the current for small modulation
amplitude is

\begin{eqnarray}
I\left(V_{0} + V_{mod}\text{sin}(\omega t)\right) = I(V_{0}) +
\left(\frac{\partial I}{\partial V}\right)_{V_{0}} \text{sin}(\omega t)
\nonumber \\
- \frac{1}{4}\left( \frac{\partial ^{2}I}{\partial V^{2}}\right)
V^{2}_{mod}\text{cos}(2\omega _{0}\tau^{\prime}) + ... . \label{expand}
\end{eqnarray}

\noindent Combining Eqs.\,\ref{apdi}, \ref{twoomega} and \ref{expand} we obtain,

\begin{eqnarray*}
&& \left( \frac{\partial ^{2}I}{\partial V^{2}}\right)_{V_{0}} = -
\int_{0}^{\infty} \frac{4}{V_{mod}^{2}} \frac{2e}{h} 2 \sum_{n=1}^{N} T_{n}
\frac{\hbar}{\tau}J_{2}(eV_{mod}\tau/\hbar)\\
& &\times \Bigl(-ir_{n}a_{l_{nn}}(\tau)\text{e}^{ieV_{0}\tau/\hbar}+
ir^{\prime}_{n}a_{r_{nn}}(\tau)\text{e}^{-ieV_{0}\tau/\hbar}+ \\
& &ir^{*}_{n}a^{*}_{l_{nn}}(\tau)\text{e}^{-ieV_{0}\tau/\hbar}- ir^{\prime *
}_{n}a^{*}_{r_{nn}}(\tau)\text{e}^{+ieV_{0}\tau/\hbar} \Bigr) d\tau.
\end{eqnarray*}

\noindent We now can continue as before and calculate the correlation function.
We assume only terms with the form $a_{l_{nn}}(\tau)a_{l_{nn}}^{*}(\tau)$ or
$a_{r_{nn}}(\tau)a_{r_{nn}}^{*}(\tau)$ contribute to this function as all other
combinations of reflection coefficients and combinations of different path times
will average to zero, and introduce the brackets which represent averaging
over different impurity configurations,

\begin{eqnarray*}
\sigma_{GV}^{2} &=& \left\langle  \left(\frac{\partial^{2} I}{\partial
V^{2}}\right)^{2}  \right\rangle =
\frac{16}{V_{mod}^{4}}\left(\frac{2e}{h}\right)^{2} 4 \\
&& \times \sum_{n=1}^{N}T_{n}^{2}(1-T_{n})
\int_{0}^{\infty} \left(\frac{\hbar}{\tau}\right)^{2}
J_{2}^{2}(eV_{mod}\tau/\hbar) \\
&& \times \langle a_{l_{nn}}(\tau)a^{*}_{l_{nn}}(\tau) +
a_{r_{nn}}(\tau)a^{*}_{r_{nn}}(\tau) \rangle d\tau.
\end{eqnarray*}

\noindent Both sides of the contact have the same average properties, and we can
write for $\langle a_{l_{nn}}(\tau)a_{l_{nn}}^{*}(\tau) \rangle= \langle
a_{r_{nn}}(\tau)a_{r_{nn}}^{*}(\tau) \rangle = P_{cl}(\tau)$. Filling in the
expression for the classical return probability (Eq.\,\ref{pcl}) and
substituting $x=eV_{mod} \tau/\hbar$ gives us,

\begin{eqnarray*}
\sigma_{GV}^{2} = \frac{16}{V_{mod}^{4}}\left(\frac{2e}{h}\right)^{2} 16
\sum_{n=1}^{N}T_{n}^{2}(1-T_{n})\\
\times \frac{\hbar^{2} v_{F}}{2 \sqrt{3 \pi} k_{F}^{2} D^{3/2}(1-\text{cos}\gamma} \left(
\frac{eV_{mod}}{\hbar}\right)^{5/2} \int_{0}^{\infty}
\frac{1}{x^{7/2}}J_{2}^{2}(x) dx.
\end{eqnarray*}

\noindent Numerical integration of the integral yields the value 0.03385 and
filling in
$D=v_{F}^{2} \tau_{e}/3$ finally results in Eq.\,\ref{sigma}.

\end{multicols}


\begin{thebibliography}{99}
\bibitem{correction} In Eq.\,2 of Ref. \protect{\onlinecite{me}} the square root over the $1-\text{cos}\gamma$ has incorrectly been left out. This error only has consequences for the elastic scattering length extracted from the experimental $\sigma_{GV}$. As will become apparent below, this $l_{e}$ is now in much better agreement with the $l_{e}$ obtained from the shift in the integer conductance values.
\end{thebibliography}
\end{document}